\documentclass{aa}  

\usepackage{graphicx}
\usepackage{txfonts}

\usepackage{hyperref}
\usepackage{fancyhdr}

\hypersetup{
    colorlinks=true,
    linkcolor=blue,
    urlcolor=blue,
    citecolor = blue,
}

\usepackage[normalem]{ulem}

\begin{document} 

   \title{Chemical characterisation of the X-shooter Spectral Library (XSL): [Mg/Fe] and [Ca/Fe] abundances\thanks{[Mg/Fe] and [Ca/Fe] values are are only available in electronic form at the CDS via anonymous ftp to \url{cdsarc.cds.unistra.fr (130.79.128.5)} or via \url{https://cdsarc.cds.unistra.fr/cgi-bin/qcat?J/A+A/}}}
   
   \author{P. Santos-Peral
          \inst{1,2}
          \and
          P. S\'{a}nchez-Bl\'{a}zquez\inst{1,2}
          \and
          A. Vazdekis\inst{3,4}
          \and
          P. A. Palicio\inst{5}
          }

\institute{Departamento de F\'{i}sica de la Tierra y Astrof\'{i}sica, Universidad Complutense de Madrid, 28040 Madrid, Spain\\
              \email{pasant02@ucm.es}
         \and
              Instituto de F\'{i}sica de Part\'{i}culas y del Cosmos IPARCOS, Facultad de CC F\'{i}sicas, Universidad Complutense de Madrid, 28040 Madrid, Spain
         \and
             Instituto de Astrof\'{i}sica de Canarias, E-38200 La Laguna, Tenerife, Spain
        \and
             Departamento de Astrof\'{i}sica, Universidad de La Laguna, E-38205, Tenerife, Spain
        \and
            Université Côte d’Azur, Observatoire de la Côte d’Azur, CNRS, Laboratoire Lagrange, France 
             }

   \date{Received December 2 2022 / Accepted February 8 2023}

 
  \abstract
   {The X-shooter Spectral Library (XSL) is a large empirical stellar library used as a benchmark for the development of stellar population models. The inclusion of $\alpha$-element abundances is crucial to disentangling the chemical evolution of any stellar system.}
   {The aim of this paper is to provide a catalogue of high-precision, accurate magnesium and calcium abundances from a wide variety of stars that are well distributed in the Hertzsprung-Russell (HR) diagram.}
   {We originally performed an analysis of the derived Mg and Ca abundances for medium-resolution spectra of 611 stars from the XSL Data Release~2. For this purpose, we used the GAUGUIN automated abundance estimation code to fit the ultraviolet-blue~(UVB) and visible~(VIS) spectra. We tested the consistency of the atmospheric parameters and chemical abundances with the \emph{Gaia} DR3 and the AMBRE Project datasets.}
   {We finally obtained precise [Mg/Fe] and [Ca/Fe] abundances for 192 and 217 stars, respectively, from which 174 stars have measurements in both elements. The stars cover a broad effective temperature range of 4000~$\textless$~T$_{\rm eff}$~$\textless$~6500~K, surface gravity of 0.3~$\textless$~log(g)~$\textless$~4.8~cm~s$^{\rm -2}$, and metallicity of  -2.5~$\textless$~[Fe/H]~$\textless$~+0.4~dex. We find an excellent agreement with the abundance estimates from the AMBRE:HARPS and the \emph{Gaia}-RVS (Radial Velocity Spectrometer) analysis. Moreover, the resulting abundances reproduce a plateau in the metal-poor regime followed by a decreasing trend even at supersolar metallicities, as predicted by Galactic chemical evolution models.}
   {This catalogue is suitable for improving the modelling of evolutionary stellar population models with empirical $\alpha$ enhancements, which could significantly contribute to the analysis of external galaxies' abundances in the near future.}


\maketitle
%

\section{Introduction}


Empirical stellar spectral libraries are key for the development of stellar population synthesis models to decode the enclosed information of unresolved stellar populations \citep[see the complete review by][]{Conroy2013}. The integrated properties of a galaxy can be modelled as a combination of single stellar populations \citep[SSP; e.g.][]{Tinsley1972, Arimoto1986, BruzualCharlot2003, Molla2009, MartinManjon2010, Vazdekis2010, GarciaVargas2013, Vazdekis2016, Maraston2020} due to the simultaneous presence of stars with different masses, ages and chemical compositions. 
Therefore, the completeness of the library in all stellar evolutionary phases, parameter space (effective temperature, surface gravity, global metallicity) and chemical abundances, is required to build more sophisticated models that will make it possible to study galaxies of varying types. \par 

Empirical libraries have the advantage of containing real observed stars, although they are limited by resolution and wavelength coverage and constrained by the presence of stars in the Galaxy to cover the Hertzsprung-Russell (HR) diagram in-depth. Recent efforts have been made to create large empirical spectral libraries with longer wavelength ranges and moderately high spectral resolution, such as the fully empirical MILES library \citep{SanchezBlazquez2006, FalconBarroso2011} on which the MILES stellar population models \citep{Vazdekis2010} and its extended version \citep[E-MILES,][]{Vazdekis2016} were based. This is also the case with a higher resolution but shorter spectral ranges such as the MEGASTAR library \citep{Carrasco2021}, which is crucial for the correct interpretation of the MEGARA multi-object spectrograph data \citep{GildePaz2018, Dullo2019} and also used for the evolutionary synthesis \texttt{POPSTAR} models \citep[see e.g.][]{Molla2009, MartinManjon2010, GarciaVargas2013}. A recent updated summary of empirical stellar libraries is listed in \citet{XshooterDR3}.  \par

In this context, the X-shooter Spectral Library\footnote{\url{http://xsl.astro.unistra.fr}} \citep[hereafter XSL, three public Data Releases up to date: DR1, DR2, DR3; ][]{XshooterDR1, XshooterDR2, XshooterDR3} is an empirical stellar spectral library of a total of 830 spectra for 683 stars, observed at a moderate-resolution (R$\sim$10000) by the X-shooter three-arm spectrograph on ESO's VLT \citep{Vernet2011}. These three spectral ranges cover UVB (ultraviolet-blue; 300-556 nm), VIS (visible; 533-1020 nm), and NIR (near-infrared; 994-2480 nm). The library aims to cover the whole HR diagram homogeneously with a wide range of chemical compositions \citep[see][which provided a uniform set of stellar atmospheric parameters for 754 spectra of 616 stars]{Xshooter_Param}. The XSL can be considered as a reference stellar library in the optical and NIR. Simple stellar population models based on the XSL were recently presented by \citet{Verro2022}. \par

Complementary high-precision chemical abundances are essential to properly analyse the star formation history (SFH) and the evolution of a stellar system. Stellar upper atmospheres provide fossil evidence of the available metals in the interstellar medium (ISM) at the time of its formation \citep{freeman2002}. From the chemical composition, it is possible to infer the properties and enrichment processes involved in the galaxy formation and evolution: star formation rate (SFR), initial mass function (IMF), production of elements (yields) for stars at different masses and timescales, or the gas evolution (migration, infall, outflows). \par 

In particular, the abundance of $\alpha$ elements (e.g. O, Mg, Si, S, Ca, Ti), relative to that of iron ([$\alpha$/Fe]), is an important fossil signature in Galactic archaeology to trace the chemical evolution. This abundance ratio is commonly used as a good chronological proxy due to the timescale delay between the rapid enrichment of the ISM with $\alpha$-elements by core-collapse supernovae (Type~II SNe, $\textless$~10$^7$~years) and the predominant release of iron-peak elements by Type~Ia SNe explosions \citep[at least 10$^8$$^-$$^9$ years; see e.g.][]{matteucci1986}. This behaviour reproduces a flat trend at low metallicities, followed by an [$\alpha$/Fe] abundance decline in the ISM. The position of the reported knee will depend on the SFH and the environment of the analysed Galaxy or any other stellar system. For instance, an active SFR will present the knee at higher metallicity regimes \citep[i.e. more contribution of massive stars;][]{matteucci1989}, and the more massive the IMF, the higher the [$\alpha$/Fe] abundance in the flat trend. In conclusion, a good coverage of stars with [$\alpha$/Fe] abundances over the stellar atmospheric parameter space will make it possible to build new updated SSPs with empirical $\alpha$-enhancements, improving the application of evolutionary stellar population synthesis methods \citep{Milone2011}. \par

Currently, there are stellar population synthesis models that predict line strengths \citep[e.g.][]{thomas2003, thomas2011, schiavon2007} and SSP spectra \citep[e.g.][]{Vazdekis2015, conroy2018} with varying abundance $\alpha$-element ratios. These models helped us to properly constrain abundance-element ratios in galaxies with unprecedented levels of detail. Previously, such studies have been done through abundance-ratio proxies based on model predictions that follow the abundance pattern of the Galaxy \citep[see][for a very detailed description of the approach]{Vazdekis2015}. These models with variable [$\alpha$/Fe] have been useful to show \citep[see][]{Vazdekis2016} that $\alpha$-enhanced populations have a flux excess blueward of $\sim$~4500 $\AA$ with respect to solar-scaled models. This is especially relevant for studies of high-redshift galaxies, where SFHs are usually constrained using a spectral energy distribution (SED) fitting with rest-frame photometric bands, but it is also very important in studies of nearby galaxies, where a full spectral fitting is commonly used to derive the SFH. The most popular approach to building up stellar populations models with varying abundance ratios is to correct observed stars with differential corrections derived from theoretical star spectra \citep[e.g.][]{knowles2021}, including the isochrones and variations in the spectra of individual stars to [$\alpha$/Fe] variations. Regarding our work, models of varying [$\alpha$/Fe] using the XSL have not been done before. The XSL represents a further step to improve these models as a result of its wide wavelength spectral coverage and resolution. For instance, it will allow us to explore the changes in the continuum in other wavelength ranges or define new indices uncontaminated from $\alpha$-element contributions (e.g. higher order Balmer lines). \par

On the other hand, Mg and Ca have distinct nucleosynthesis sites (magnesium is produced almost entirely in core-collapse supernovae, while some calcium is also produced in exploding white dwarfs), and the comparison between both can provide relevant information to understand the evolution of asymptotic giant branch (AGB) stars, the IMF (production of Mg/Ca in core-collapse supernovae varies with the mass of the progenitor), or the SFH. For instance, new calcium-rich type II SNe have been found by several authors \citep[e.g.][]{Perets2010, De2020}, predominantly in early-type galaxies and probably due to the collapse of AGB stars, which could significantly contribute to the Ca production. Furthermore, the different production sites of Ca and Mg are also evident when we compare the behaviour of  [Ca/Fe] and [Mg/Fe] with velocity dispersion in elliptical galaxies. Several authors have shown, using different Ca-sensitive lines in the optical and the NIR, that for increasing velocity dispersions, [Mg/Fe] increases steeply, while [Ca/Fe] decreases \citep[e.g.][]{Vazdekis1997, Cenarro2003, Cenarro2004}. \par

The main objective of this work is to provide reliable [$\alpha$/Fe] abundances for one of the most complete empirical stellar libraries in the literature. In this paper, we present a detailed spectroscopic analysis of the Mg and Ca abundance estimation for a sample of 611 stars observed and parametrised at medium spectral resolution within the context of the X-shooter Spectral Library \citep{Xshooter_Param, XshooterDR2}. The determined abundance ratios are required for building up models with varying Mg and Ca abundances, which differential corrections can be used to correct the XSL stars to compute semi-empirical spectra and, consequently, the stellar population models that implement these stars. \par

The paper is organised as follows. In Sect. \ref{data}, we introduce the observational data sample used in the present analysis. The automatic abundance estimation method is described in Sect. \ref{method}. In Sect. \ref{results}, we show the final derived [Mg/Fe] and [Ca/Fe] abundances of the sample. We conclude with a summary in Sect. \ref{conclusions}.


\section{The X-shooter observational data sample} \label{data}

The XSL observations were obtained during six semesters by the X-shooter three-arm spectrograph at the ESO's VLT telescope \citep{Vernet2011}. They comprise a large variety of stars covering the HR diagram, where more than a half are giant stars, with a wide range of metallicities \citep[2900~$\textless$~T$_{\rm eff}$~$\textless$~38000~K, 0~$\textless$~log(g)~$\textless$~5.7~cm~s$^{\rm -2}$, and -2.5~$\textless$~{[}Fe/H{]}~$\textless$~+1.0~dex; we refer the reader to Fig.~7 or Fig.~1 in][respectively]{Xshooter_Param, XshooterDR3}. The target list of stars in the XSL sample were configured from existing spectral libraries and the parameter compilation catalogue PASTEL \citep{soubiran2010, soubiran2016}. A detailed table with the used catalogues is presented in \citet{XshooterDR2}. This approach gives a strong overlap with other observational spectral libraries such as MILES \citep[142 stars,][]{SanchezBlazquez2006, FalconBarroso2011}, NGSL \citep[135 stars,][]{gregg2006}, and ELODIE \citep[][]{prugniel2001, prugniel2004, prugniel2007}.  \par

As mentioned above, the X-shooter instrument splits the light in three-arm spectrographs that cover different spectral wavelength domains and mean resolving powers: UVB (300-556~nm; R$\sim$10000), VIS (533-1020~nm; R$\sim$11000), and NIR (994-2480~nm; R$\sim$8000). In the present paper, we analyse the UVB and VIS arms from the XSL Data Release~2 \citep{XshooterDR2} separately; this contains 813~spectra of 666~stars, including the re-reduced spectra of DR1 \citep{XshooterDR1}. Typical signal-to-noise ratios (S/N) are about 70 and 90 for the UVB and VIS arms, respectively, except for coolest stars. The DR2 data were homogeneously reduced and calibrated, corrected for atmospheric extinction, telluric absorption and also by radial velocity (i.e. spectra in the rest frame, wavelengths in air). Furthermore, stellar atmospheric parameters (T$_{\rm eff}$, log(g), [Fe/H]) for 754 spectra (from UVB and VIS arms) of 616 XSL DR2 stars were estimated by \citet{Xshooter_Param}, who also performed an exhaustive comparison with the literature. The averaged internal uncertainties in T$_{\rm eff}$, log(g), [Fe/H] are $\sim$~80K, 0.15~dex, and 0.09~dex, respectively. Recently, the XSL DR3 was published \citep{XshooterDR3}, extending the sample with 20 M dwarfs, which are outside the scope of this paper due to their low temperature values. \par

We note that the XSL DR2 dataset contains some repeated observations for some stars. For these cases, we used the known \textit{Gaia} DR2 ID \citep{gaia2018} of each spectrum to identify the different spectra of the same star, and then we estimated the median stellar atmospheric parameters. Moreover, we excluded the individual spectra whose atmospheric parameters differed by more than two sigma from the median value of the spectrum distribution per star. This rejection could help to avoid the propagation of uncertainties on the atmospheric parameters to the stellar abundances and discard possible spectroscopic binaries. This selection based on the known parameters from \citet{Xshooter_Param} lead to a total of 748 spectra, corresponding to 611 stars.

\subsubsection*{Parameter validation} \label{Parameter_Validation}

As mentioned above, the stellar atmospheric parameters (T$_{\rm eff}$, log(g), [Fe/H]) for the XSL DR2 sample were previously analysed and determined by \citet{Xshooter_Param}. In brief, they employed the full-spectrum fitting package, ULySS \citep{Koleva2009}, using both the UVB and visible VIS spectra and the MILES empirical spectral library \citep{SanchezBlazquez2006, FalconBarroso2011} as references. This package performs a local minimisation at each wavelength bin between the observed spectrum and an interpolated one from the reference stellar spectral library for the given stellar atmospheric parameters T$_{\rm eff}$, log(g), and [Fe/H] \citep[computed with the MILES spectral interpolator;][]{Prugniel2011, sharma2016}, using the whole spectral range without a strong influence of the spectral resolution \citep[see e.g. ][]{Prugniel2011, KolevaVazdekis2012}. It is worth mentioning that the MILES parameters were not measured directly on the MILES spectra, but are bound to literature compilations \citep[e.g. PASTEL catalogue;][]{soubiran2016} and homogenised as in \citet{cenarro2007}. Only measurements made on  high resolution spectra and with a low uncertainty were included. Additionally, they carried out an extensive comparison with measurements in the literature (values obtained using high-resolution spectra) from PASTEL, finding a great level of consistency, with an agreement within 2.7\%, 0.36, and 0.15 for T$_{\rm eff}$, log(g), and [Fe/H]. \par

\begin{figure*}
\centering
\hspace*{-0.7cm}
\includegraphics[height=150mm, width=\textwidth]{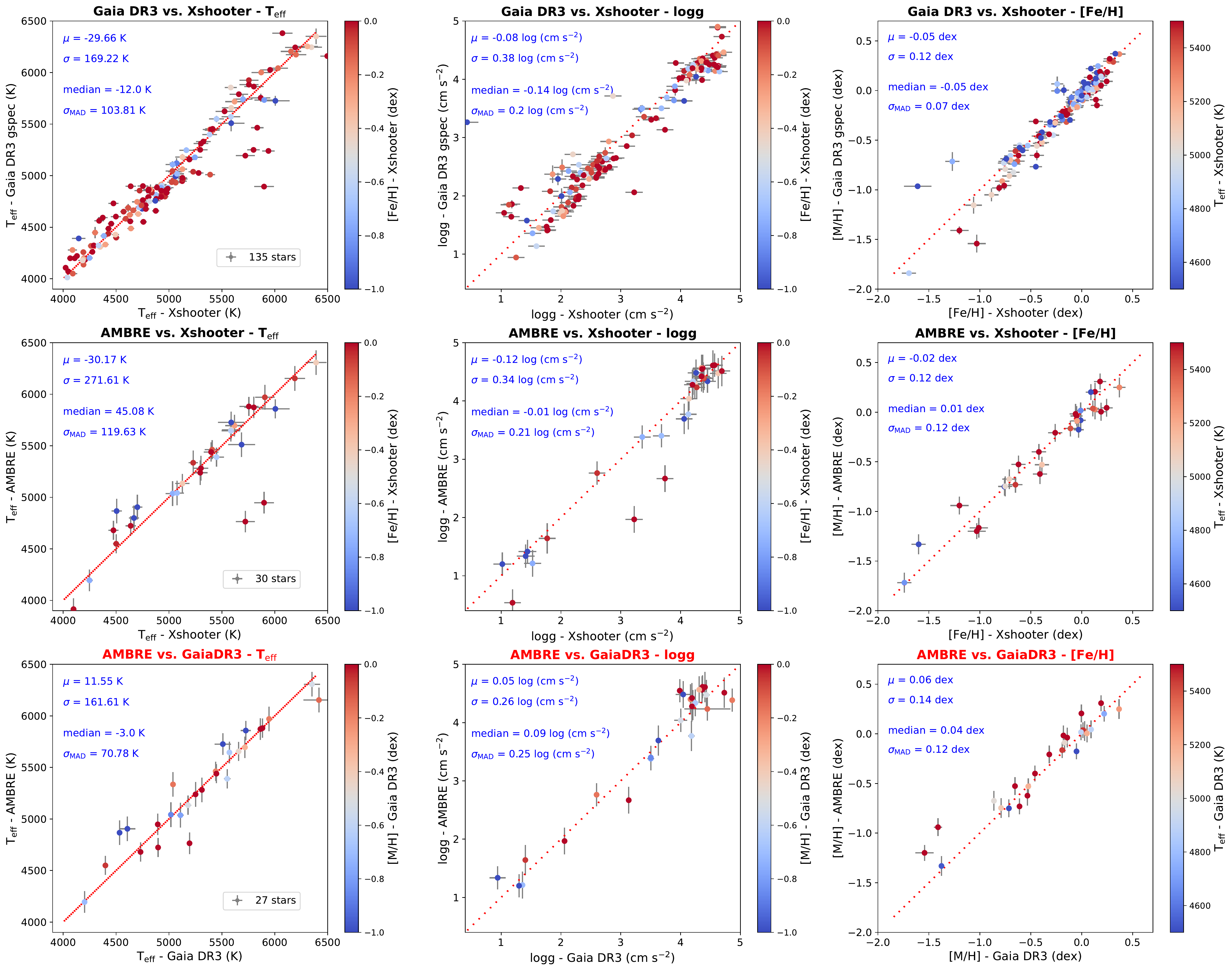}
\caption{Comparison between used X-shooter stellar atmospheric parameters in this work and some literature data. Each atmospheric parameter is shown in a separate column: effective temperature (\textit{left}), surface gravity (\textit{middle}), and metallicity (\textit{right}). The first two columns are colour-coded by the stellar metallicity, while the last one is colour-coded by the effective temperature. \textit{Top row}: \emph{Gaia} DR3 vs. X-shooter. \textit{Middle row}: AMBRE data vs. X-shooter. \textit{Bottom row}: AMBRE vs. \emph{Gaia} DR3 for the stars in common with the X-shooter spectral library. The mean ($\mu$), standard deviation ($\sigma$), median, and robust standard deviation (i.e. $\sim$~1.48 times the median absolute deviation, MAD) of the offsets are indicated in each panel.}
\label{Fig:Parameter_comparison}
\end{figure*}

Beyond the extensive analysis done by \citet{Xshooter_Param} on the derived stellar atmospheric parameters, we performed additional tests for the validation of the implemented parameters with more recent published datasets: \emph{Gaia\emph{'s}} third data release~(DR3) \citep[][]{GaiaDR3} and the AMBRE Project \citep[][]{patrick2013}. In this work, we implemented the same methodology as previously done with AMBRE \citep[][]{SantosPeral2020} and \emph{Gaia} DR3 data \citep[][]{alejandra2022_RVS} for deriving individual chemical abundances (fully described in Sect.~\ref{method}). Therefore, these catalogues can be used as a reference for the comparison of our abundance estimates from XSL observed spectra. For that purpose, in order to prevent potential differences that could arise from the used input stellar parameters from the XSL DR2, we performed a comparison analysis of the stellar atmospheric parameters with those stars in common with AMBRE and \emph{Gaia} stellar samples.\par

First, the XSL sample presents a great overlap (606~stars) with the most recent \emph{Gaia} Radial Velocity Spectrometer (RVS) data catalogue, parametrised by the General Stellar Parametriser-spectroscopy (\emph{GSP-Spec}) module \citep[][]{alejandra2022_RVS}. Therefore, we cross-matched them, only taking into account their stellar atmospheric parameters that followed the best quality criteria defined in \citet{alejandra2022_RVS}: only stars that have their first 13 quality flags equal to zero (see their Table~2), except their \emph{extrapol} flag (which is indicative of the extrapolated results near the grid limits), for which we allowed it to be less than or equal to one (A.~Recio-Blanco and P.A. Palicio, priv. comm.). We found a total of 155 stars in common with the X-shooter library. In addition, we applied the calibration suggested by \citet{alejandra2022_RVS} on the \emph{Gaia} DR3 parameters to correct them from the reported global biases with respect to the literature. 
We properly corrected the uncalibrated gravities (slightly biased in $\sim$~0.3~dex, larger for giants than for dwarfs) and metallicities (with an overall zero offset, but slightly underestimated in giants and overestimated in dwarfs), by fitting the recommended low-order polynomial \citep[coefficients are shown in Table 3 of][]{alejandra2022_RVS} as a function of their uncalibrated surface gravity. \par

Secondly, we compared the XSL DR2 parameters with those stars in common with the AMBRE Project: a total of 32 stars. The AMBRE \textbf{P}roject is a collaboration between the Observatoire de la Côte d’Azur (OCA) and the European Southern Observatory (ESO) to automatically and homogeneously parametrise archived stellar spectra from ESO spectrographs: FEROS (R~=~48000), HARPS (R~=~115000) and UVES (R~$\sim$~47000). The AMBRE parametrisation analysis for each ESO spectrograph database are fully described in \citet[][FEROS and UVES, respectively]{worley2012, worley2016} and \citet[][HARPS]{DePascale2014}. The stellar atmospheric parameters from both the \emph{Gaia} DR3 and the AMBRE data were similarly derived by the multi-linear regression algorithm MATISSE \citep[MATrix Inversion for Spectrum SynthEsis,][]{alejandra2006}, which was also used in the \emph{Gaia} RVS analysis \citep[R$\sim$11500,][]{alejandra2016, alejandra2022_RVS}. MATISSE is a projection method that derives stellar parameters by the projection of the input-observed spectrum on a set of vectors, which represent a linear combination of reference synthetic spectra \citep[see a detailed description in][]{worley2012}. \par

The comparison of the stellar parameters among the three discussed catalogues is shown in Fig.~\ref{Fig:Parameter_comparison}, with their respective associated errors. We only plot stars with 4000~$\textless$~T$_{\rm eff}$~$\textless$~6500~K, since reliable chemical abundances could not be derived out of this regime (described in Sect.~\ref{flags_and_errors}). We observe a very good agreement between datasets, with small biases for each atmospheric parameter. For our XSL sample, we find a median offset for T$_{\rm eff}$, log(g), [Fe/H] of -12~K, -0.14~cms$^{-2}$ and -0.05~dex with respect to the \emph{Gaia} DR3 catalogue, and 45~K, -0.01~cms$^{-2}$ and 0.01~dex compared with the AMBRE Project dataset; this shows a robust standard deviation of 104~K, 0.2~cms$^{-2}$, and 0.07~dex, and 119~K, 0.21~cms$^{-2}$, and 0.12~dex, respectively. In particular, we only observe a few hot XSL stars (T$_{\rm eff}$~$\textgreater$~5500~K) with larger differences in T$_{\rm eff}$ with respect to the \emph{Gaia} and AMBRE estimates. Regarding the surface gravity and metallicity comparison, we show slightly more dispersed values for giant (log(g)~$\lesssim$~3.5~cms$^{-2}$) and metal-poor stars ([Fe/H]~$\lesssim$~-1.0~dex). When comparing the \emph{Gaia} DR3 and AMBRE catalogues, we obtain similar trends for the stars in common with the X-shooter observational data sample. It is worth noting that the [M/H] notation from the \emph{Gaia} DR3 and the AMBRE catalogues is actually tracing the [Fe/H] abundance (later described in Sect.~\ref{Spectra_grids} for the computation of the synthetic spectra). In conclusion, based on the very good agreement found among the discussed datasets, we are able to re-confirm the reliability of the stellar atmospheric parameters determined by \citet{Xshooter_Param} in the analysed range in effective temperature in this work: 4000~$\textless$~T$_{\rm eff}$~$\textless$~6500~K.


\section{Method} \label{method}
From the medium-resolution observational spectra sample described in the previous section, we derived and analysed the [Mg/Fe] and [Ca/Fe] abundances from individual spectral lines in the optical and NIR range via the automated abundance estimation code GAUGUIN \citep{Bijaoui2012, alejandra2016}. GAUGUIN is an optimised spectrum synthesis algorithm for deriving precise abundances, developed in the framework of the \emph{Gaia}-RVS analysis pipeline \citep[][]{alejandra2016, alejandra2022_RVS} and applied within the AMBRE Project context \citep{guiglion2016} and the Gaia-ESO Survey pipeline \citep{alejandra2014}. \par

Globally, the determination of high-precision abundances is constrained by the need for both high S/N and spectral resolution, and predominantly by the definition of continuum to normalise the observed spectral data \citep{nissen2018, BIBLIA_Jofre}. An in-depth analysis of how these issues affect the applied methodology in this work can be found in \citet{SantosPeral2020}, who also showed a significant improvement in the precision of abundance estimates by carrying out an optimisation of the spectral normalisation procedure. In brief, the normalisation procedure has an intrinsic dependence on the stellar type and the intensity of the analysed line that can be reduced by exploring the application of narrow normalisation windows, which additionally reduce the line-to-line abundance scatter.

\subsection{\textbf{Adopted line list}} \label{line_list}

For the selection of our analysed lines, we referred to the works of \citet{SantosPeral2020}, who performed an extensive spectroscopic analysis of nine different magnesium lines in the optical range, and \citet{contursi2021} and \citet{alejandra2022_RVS}, who compiled the spectral analysis of \emph{Gaia}~DR3, the wavelength interval domain of which was chosen around the Ca II IR triplet with a similar spectral resolution of R$\sim$11500. In the present analysis, we selected strong well-known lines in order to ensure that its abundance can be properly measured from any stellar type (even the most metal-poor ones, i.e. [Fe/H] < -2.0~dex) and at low-medium spectral resolution. \par

The abundance analysis was performed using three magnesium spectral lines in the optical range and three near-infrared calcium lines. They are shown in Table \ref{table:lines} and illustrated later in Fig.\ref{Fig:MgCa_windows}: \par

\begin{table}[h]
\centering
\begin{tabular}{ccc}
\hline
\hline
\multicolumn{3}{c}{\textbf{Mg I (\AA):}} \vspace{0.05cm} \\
5167.3 & 5172.7 & 5183.6 \\
\hline
\multicolumn{3}{c}{\textbf{Ca II (\AA):}} \vspace{0.05cm} \\
8498.02 & 8542.09 & 8662.14 \\
\hline
\hline
\end{tabular}
\vspace{0.08cm}
\caption{Magnesium and calcium lines (vacuum wavelengths) selected in the present analysis.}
\label{table:lines}
\end{table}

The selected Mg Ib triplet lines are almost unaffected by non-local thermodynamic equilibrium (NLTE) effects, even showing a robust behaviour with respect to 3D NLTE calculations, but no significant differences compared to 1D LTE models \citep{bergemann2014,bergemann2017, zhao2016, alexeeva2018}. However, for the Ca II triplet it is necessary to adopt more realistic assumptions of the radiative transfer in the stellar atmospheres to properly fit both the core and the wings, since the cores are probably subject to NLTE and/or chromospheric effects \citep{contursi2021}. To minimise NLTE effects, two abundance windows were defined at the calcium line wings, avoiding the cores \citep[i.e. up to six independent abundance estimates can be provided for the three Ca II lines, as described in][]{alejandra2022_RVS}. \par

\begin{figure*}[h]
\centering
\includegraphics[width=0.8\textwidth]{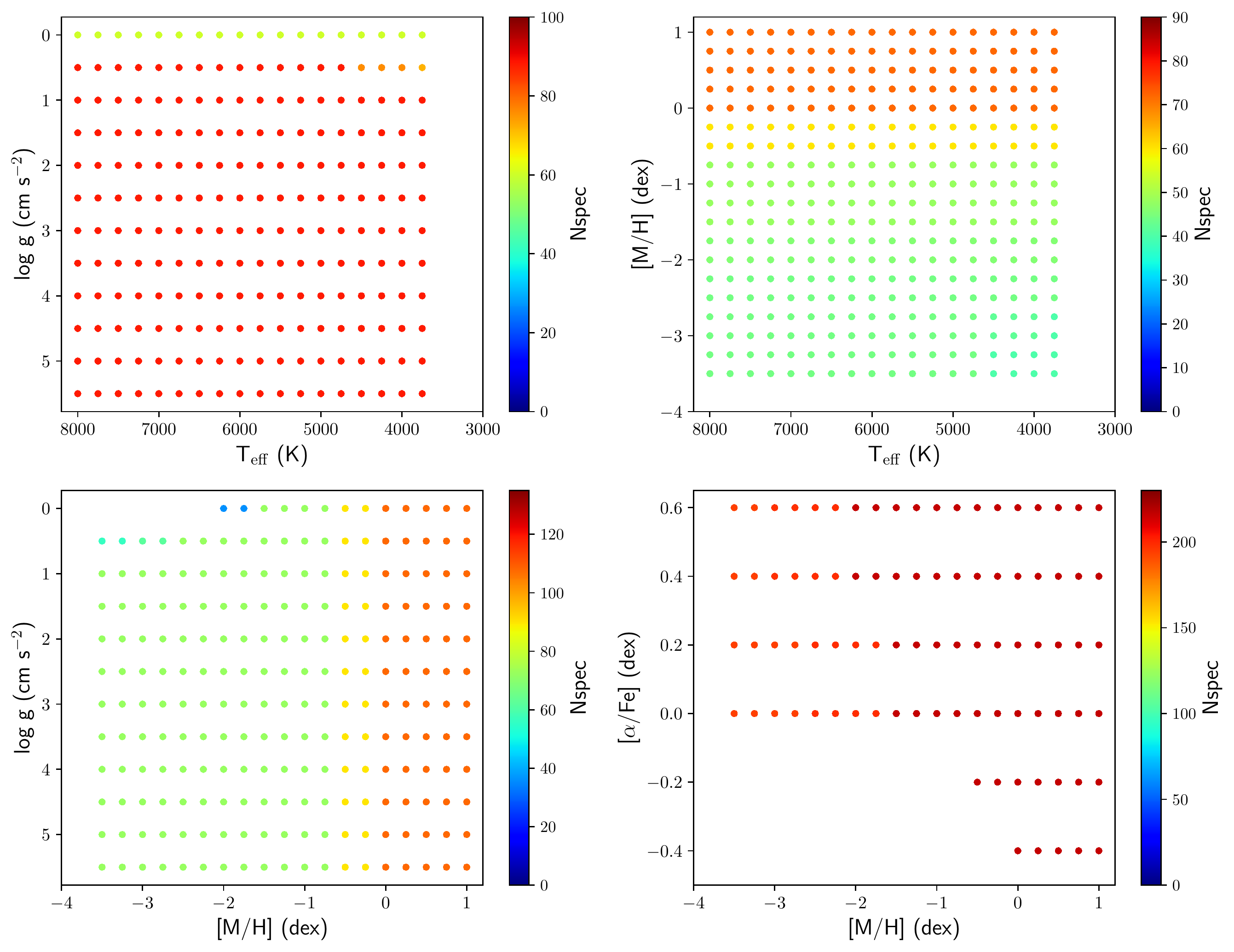} 
\caption{Stellar atmospheric parameters of reference synthetic spectra grid employed in the GAUGUIN procedure for the [Mg/Fe] abundance determination, where Nspec indicates the number of spectra at each parameter combination.}
\label{fig:GridGES_observations}
\end{figure*}

Although the selected Mg I and Ca II spectral lines are commonly used in spectroscopic analysis due to their strength, and they are known to be less affected by blends in very crowded spectra (e.g. cool, metal-rich stars) in comparison with weaker lines, it is worth noting that a small effect from blends in the continuum placement, and therefore the abundance estimate cannot completely discarded depending on the line. As discussed in detail in Sections~\ref{magnesium} and~\ref{calcium} with regard to our abundance results, the first Mg I 5167.3~$\AA$ line seems to suffer from blends with the strong Fe I 5168.89~$\AA$ line \citep[also seen in previous works including][]{aldenius2007}, and the left Ca II 8498.02~$\AA$ line could also be affected by the presence of strong titanium and iron lines \citep[already discussed in previous studies of the Ca II triplet region; e.g.][]{mallik1997}. As we show in Sect.~\ref{results} and with the biases illustrated in the appendix, both lines were eventually rejected in the final abundance analysis. \par

\subsection{\textbf{Reference synthetic spectra grids}} \label{Spectra_grids}

For the selected Mg Ib triplet lines, we followed the same methodology as \citet{SantosPeral2020} for the determination of magnesium abundances. We adopted the same 4D spectra grid in T$_{\rm eff}$, log(g), [M/H], and [$\alpha$/Fe]. It consists of a high-resolution, optical, synthetic grid (4200-6900\AA; R$\sim$300000, 18452 spectra) of non-rotating FGKM-type spectra. It was computed with the spectrum synthesis code TURBOSPECTRUM \citep{turbospectrum, Plez2012} by adopting 1D LTE MARCS atmosphere models \citep{MARCS}, solar chemical abundances of \citet{grevesse2007}, and the Gaia-ESO Survey atomic and molecular line lists \citep[][]{heiter2015,heiter2021}. The covered atmospheric parameter ranges are shown in Fig.~\ref{fig:GridGES_observations}: 3750 $\leq$ T$_{\rm eff}$ $\leq$ 8000 K (in steps of 250 K), 0.0 $\leq$ log(g) $\leq$~5.5~cm~s$^{-2}$ (in steps of 0.5 cm s$^{-2}$), and -3.5 $\leq$ [M/H] $\leq$ +1 dex (in steps of 0.25 dex), whereas the variation in [$\alpha$/Fe] is -0.4 $\leq$ [$\alpha$/Fe] $\leq$ 0.6 dex (for [M/H] $\geq$ 0.0 dex), -0.2 $\leq$ [$\alpha$/Fe] $\leq$ 0.6 dex (for -0.5 $\leq$ [M/H] $\textless$ 0.0 dex), and 0.0 $\leq$ [$\alpha$/Fe] $\leq$ 0.6 (for [M/H] $\textless$ -0.5 dex), with steps of 0.2 dex. \par

Conversely, for the derivation of calcium abundances, we used the computed sets of 5D grids implemented in the recent \emph{Gaia}-RVS analysis by the General Stellar Parametriser-spectroscopic (\emph{GSP-Spec}) module \citep{alejandra2022_RVS}, where the GAUGUIN procedure was also applied for individual chemical abundance estimations. For this grid, the first four dimensions (T$_{\rm eff}$, log(g), [M/H] and [$\alpha$/Fe]) are the ones of the 4D \emph{GSP-Spec} reference grid \citep[see the covered parameter space in Fig.~3 of][]{alejandra2022_RVS} containing 51373 synthetic spectra, calculated in the same way as described in previous paragraph. The fifth dimension corresponds to the abundance values of the specific chemical element [Ca/Fe]. It covers the wavelength range from 846 to 870~nm of the RVS spectra with a wavelength step of 0.03~nm. The variations in the chemical element dimension are from -2.0 to +2.0 dex around $\epsilon$(X) = $\epsilon$(X)+[M/H]+K$_{\alpha}$, with a step of 0.2 dex (i.e. 21 different abundance values). The variable K$_{\alpha}$ follows a similar variation with the metallicity to [$\alpha$/Fe]: K$_{\alpha}$=0.0 for [M/H]$\geq$~0.0~dex, K$_{\alpha}$=+0.4 for [M/H]~$\leq$-~1.0~dex, and K$_{\alpha}$~=-0.4 × [M/H] for -1.0 $\leq$~[M/H]~$\leq$~0.0~dex. In total, the 5D grid contains $\sim$590750 spectra covering the whole metallicity regime of the atmosphere model grids. \par

We refer to a detailed description of the synthetic spectra calculation in \citet{patrick2012}, implemented for the AMBRE Project. However, it is worth noting that the grids used in this work have been updated from that time through a more consistent [$\alpha$/Fe] enrichment considered in the model atmosphere and the calculated emerging spectra, and a more realistic approach for the microturbulent velocity value by using an empirical law parametrised as a function of the main atmospheric parameters (M. Bergemann, in preparation). No stellar rotation or macroturbulence broadening were included in these spectra. Finally, to be in agreement with the X-shooter observational spectra dataset described in Sect.~\ref{data}, we convolved with a Gaussian kernel and re-sampled the synthetic spectra to the observed resolving power (R$\sim$10000 for the UVB, and R$\sim$11000 for the VIS arms). \par

It is also important to remark that the [M/H] notation of the synthetic grids comes from the atmosphere models. However, in the computation of the synthetic spectra (where the models are used) we imposed that the [M/H] parameter must be equal to the iron abundance value ([Fe/H]) of the star. Therefore, in the following comparison with the observed X-shooter spectra, we can assume that [M/H] actually behaves as [Fe/H] since it does not take into account the $\alpha$-element enrichment.

\subsection{The GAUGUIN automated abundance estimation code} \label{GAUGUIN}

The GAUGUIN algorithm \citep{Bijaoui2012, alejandra2016, alejandra2022_RVS} is a classical local optimisation method, based on a local linearisation around an input set of parameters associated with a reference synthetic spectrum. As described in detail in \citet{SantosPeral2020}, the observed spectrum flux is normalised over a given wavelength interval around the analysed line, conserving its original spectral resolution. Then, for each line, a new specific-reference synthetic spectra grid is generated in the abundance space ([X/Fe]$_{i}$) by interpolating the input stellar atmospheric parameters (T$_{\rm eff}$, log(g), [Fe/H], [$\alpha$/Fe]). The four input atmospheric parameters of each observed spectrum are fixed and are necessary to initialise GAUGUIN to derive individual chemical abundances. In this work, the first three stellar parameters (T$_{\rm eff}$, log(g), [Fe/H]) were independently determined by \citet{Xshooter_Param}, and for the [$\alpha$/Fe] parameter we adopted an approach based on the spectrum's metallicity. \par

\begin{figure*}[h]
\centering
\hspace*{-1.2cm}
\includegraphics[height=70mm, width=0.5\textwidth]{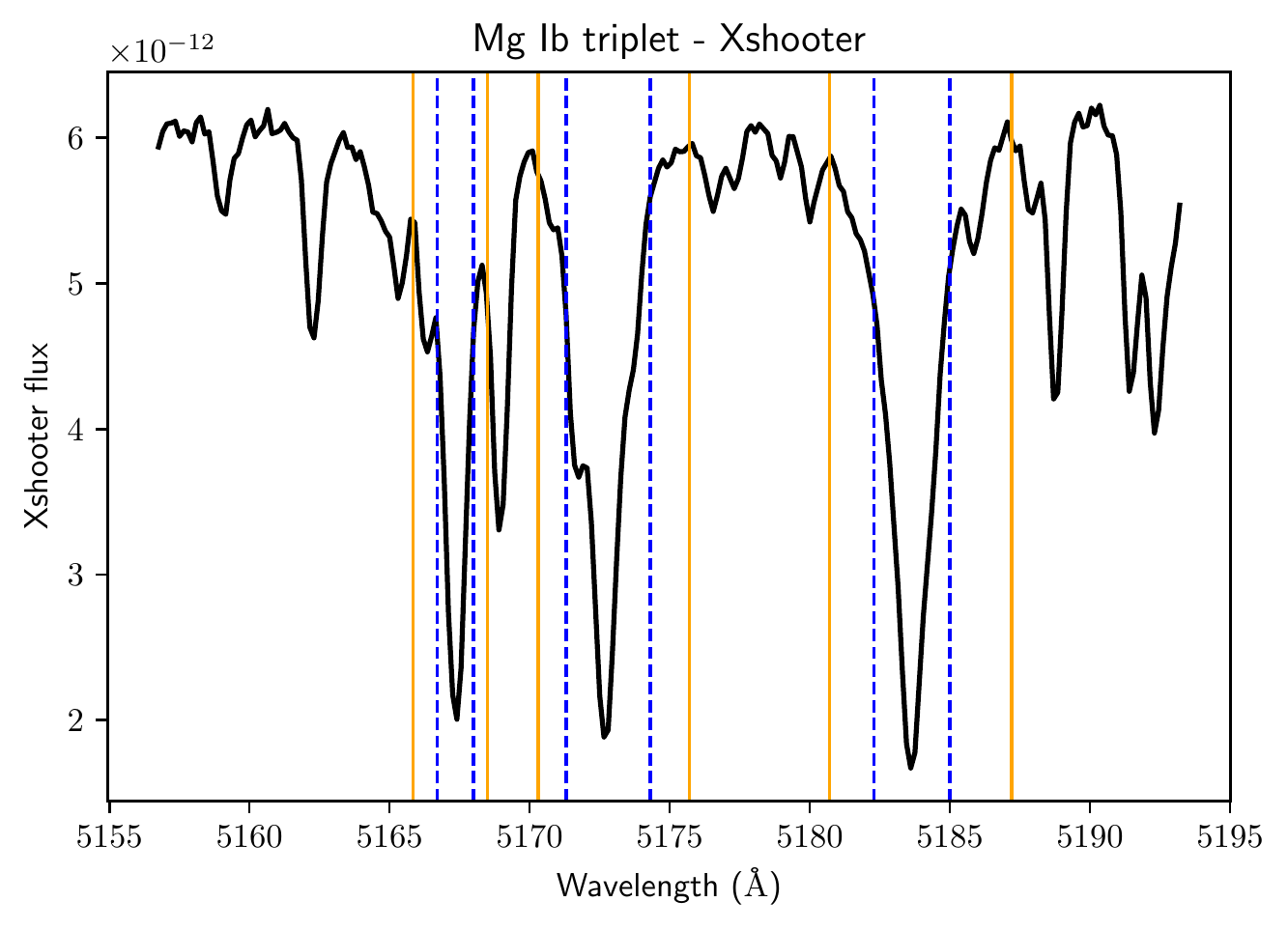}
\includegraphics[height=70mm, width=0.5\textwidth]{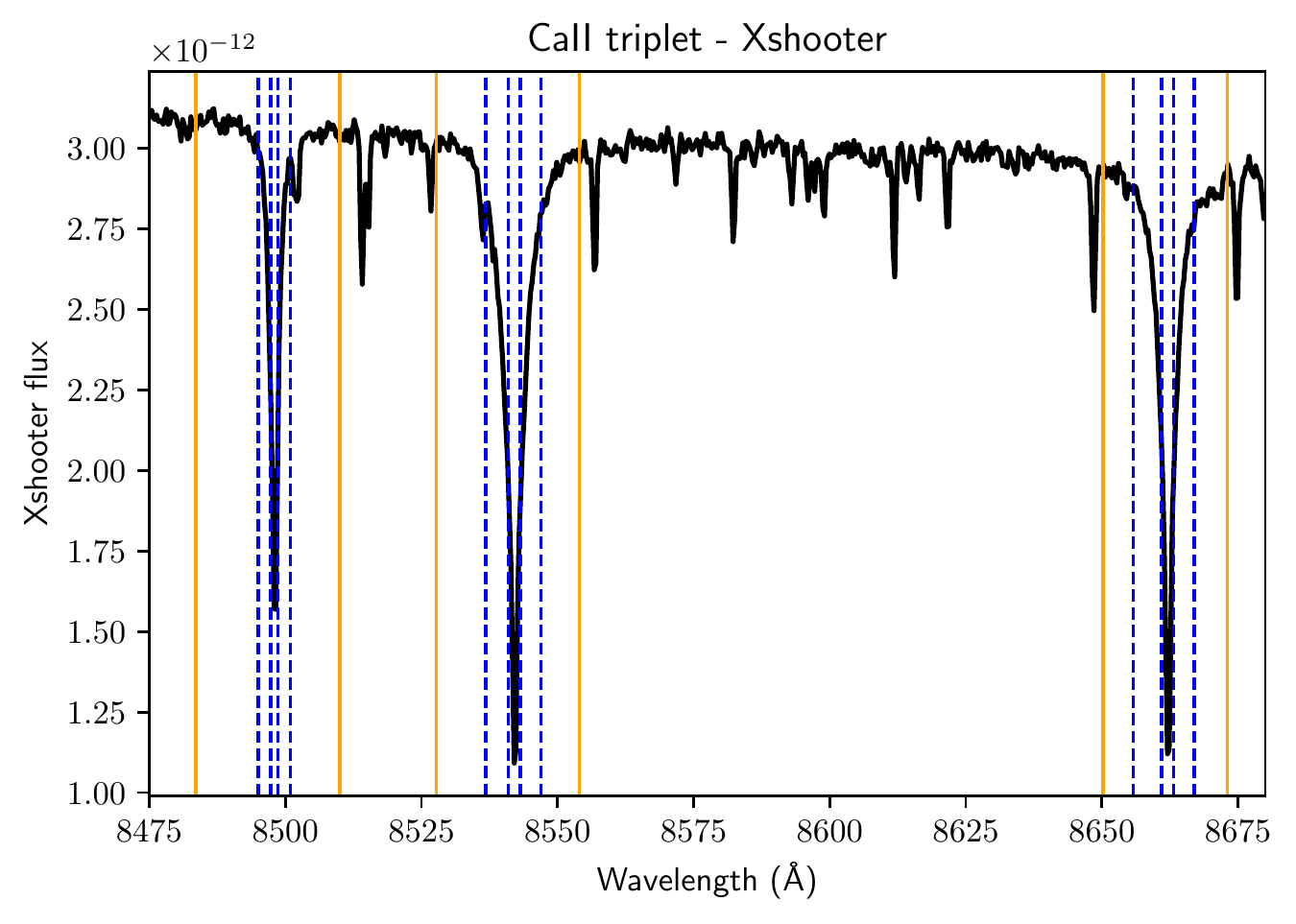}
\caption{Observed spectrum of a solar-type star from X-shooter spectrograph (R$\sim$10000) around the Mg Ib triplet lines (5167.3, 5172.7 and 5183.6~$\AA$, \emph{left panel}; S/N~$\sim$~79) and the Ca II IR triplet (8498.02, 8542.09 and 8662.14~$\AA$, \emph{right panel}; S/N~$\sim$~96). The adopted wavelength domain where the abundance is measured is delimited by blue dashed vertical lines; we note that these intervals exclude the core of the Ca II lines. An example of the adopted local normalisation intervals is shown with an orange vertical line next to each individual line, following the optimised criteria from \citet{SantosPeral2020}.}
\label{Fig:MgCa_windows}
\end{figure*}

\begin{figure*}[h]
\centering
\includegraphics[height=99mm, width=0.46\textwidth]{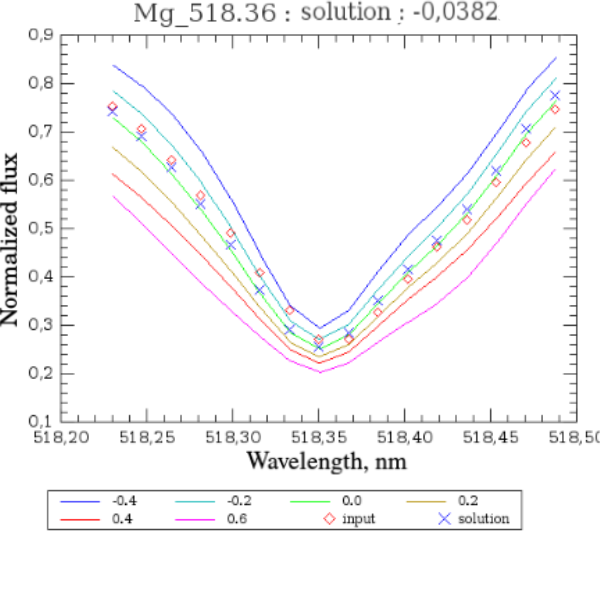}
\includegraphics[height=100mm, width=0.45\textwidth]{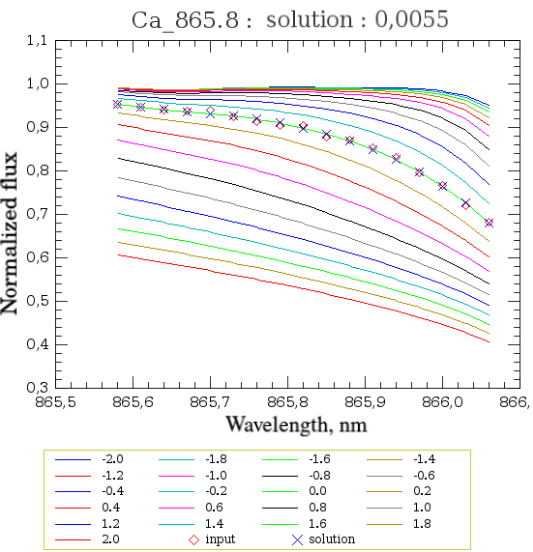}
\caption{Example of fit carried out by optimisation code GAUGUIN for Mg line 5183.6~$\AA$ (\textit{left panel}) and the left wing of the Ca line 8662.14~$\AA$ (\textit{right panel}) in a solar-type star. The normalised observed spectrum is shown with red open diamonds, while the solution is indicated by blue crosses. The reference synthetic spectra grid is colour-coded according to [$\alpha$/Fe] (\textit{left panel}) and [Ca/Fe] values (\textit{right panel}).}
\label{Fig:GAUGUIN_fit_example}
\end{figure*}

The abundance estimate is performed considering the spectral flux in a pre-defined wavelength window, which is always contained by the previously defined local normalisation one. Both intervals are not necessarily symmetric around the analysed line, and they were chosen in order to avoid contiguous strong absorption lines and maximise the number of pixels close to the continuum level (see an example in Fig.~\ref{Fig:MgCa_windows}). For the $\alpha$-element abundance determination, the synthetic grid covers different [$\alpha$/Fe] values (in case a 5D specific chemical element grid was not computed). First, a minimum quadratic distance is calculated between the reference grid and the observed spectrum, providing a first guess of the abundance estimate ([X/Fe]$_{0}$). This initial guess is then optimised via a Gauss-Newton algorithm that stops when the relative difference between two consecutive iterations ($\Delta$[X/Fe]$_{i}$) is less than a given value (one-hundredth of the synthetic grid abundance step) and provides the final abundance estimation of each line ([X/Fe]). Figure~\ref{Fig:GAUGUIN_fit_example} illustrates, for the X-shooter observed spectrum of a solar-type star, the fit carried out by the automated code GAUGUIN for the magnesium line 5183.6~$\AA$ and the calcium line 8662.14~$\AA$. \par

Regarding the [$\alpha$/Fe] abundance dimension of the XSL spectra, which is a mandatory input requirement to initialise GAUGUIN for deriving abundances and was not estimated by \citet{Xshooter_Param}, we followed two different methods: (i) the adoption of the [$\alpha$/Fe]-enrichment values from the recently published \emph{Gaia} DR3 catalogue \citep{alejandra2022_RVS, GaiaDR3} when possible; or (ii) an empirical estimation based on the [$\alpha$/Fe] versus metallicity trends observed in the Milky Way: [$\alpha$/Fe]~=~+0.4 for [M/H]~$\leq$~-1.0~dex, [$\alpha$/Fe]~=~0.0 for [M/H]$\geq$~0.0~dex and [$\alpha$/Fe]~=~-0.4 × [M/H] for -1.0 $\leq$~[M/H]~$\leq$~0.0~dex. The adopted empirical approach was inferred as the one that better reproduces the derived [Mg/Fe] abundances with GAUGUIN from the whole AMBRE:HARPS stellar sample \citep{SantosPeral2020}.

First, we followed the same cross-match criteria for the \emph{Gaia} DR3 catalogue as that described for the parameter comparison in Sect.~\ref{Parameter_Validation}. We remark that the [$\alpha$/Fe] provided in \emph{Gaia} DR3 is dominated by the Ca II Triplet lines in the RVS domain (the same atomic lines analysed in this work, see Table~\ref{table:lines}). The final quality sample of 155 stars in common with the X-shooter library mostly comprises the metal-rich regime ([Fe/H]~$\gtrsim$~-0.5~dex), from which we adopted the [$\alpha$/Fe]$_{\rm RVS}$ measurement after careful calibration. \par

Secondly, it is worth mention that the [$\alpha$-element/Fe] versus [Fe/H] plane is predicted by chemical evolution models to describe a decreasing trend even at supersolar metallicities \citep[{[}Fe/H{]}~$\textgreater$~0.0~dex, e.g.][]{romano2010, kobayashi2020, palla2020}. This behaviour has also been reported in recent observational studies for $\alpha$ element such as oxygen \citep{bensby2014}, magnesium \citep{SantosPeral2020}, sulfur \citep{perdigon2021}, and calcium \citep{alejandra2022_cartography}. However, there is not a general consensus since previous studies observed a flattened trend at supersolar metallicities \citep[see e.g.][]{vardan2012, anders2014, sarunas2017,fuhrmann2017, buder2019}. \par

The sensitivity of the final abundance estimate to the initial [$\alpha$/Fe] guess was analysed in depth in \citet{SantosPeral2020}. For stellar types with no pixels reaching the continuum level (e.g. cool and metal-rich stars), the adopted input [$\alpha$/Fe] parameter by GAUGUIN for the interpolation and normalisation procedure, assumed as a first guess for the [Mg/Fe] or any other $\alpha$-element abundance, has a strong influence on the pseudo-continuum placement. Therefore, for these particular cases, the first-guess abundance could have a significant influence on the final derived abundance value. This influence is weaker for the derivation of [Ca/Fe] abundances since we used 5D computed grids (see Sect.~\ref{Spectra_grids}), varying the calcium abundance dimension and therefore being more independent of the global [$\alpha$/Fe] spectral value. \par

In this work, the two different adopted sets of [$\alpha$/Fe] parameters for the X-shooter sample are perfectly complementary. Luckily, we have the [$\alpha$/Fe]$_{\rm RVS}$ estimates for these particular metal-rich cases with no continuum, whose reported values already took into account this effect in the GAUGUIN procedure \citep{alejandra2022_RVS}. On the contrary, for metal-poor stars, the normalisation procedure is independent of the initial [$\alpha$/Fe], and we tested that the assumed empirical approach allowed us to derive accurate and independent individual chemical abundances. In addition, as mentioned before, the observed X-shooter spectra catalogues are already corrected by their measured radial velocity, allowing direct processing of the data without any previous manipulation.

\subsection{Construction of the X-shooter catalogue} \label{flags_and_errors}

Based on the line selection shown in Table~\ref{table:lines}, we analysed the observed X-shooter spectra from the UVB (300-556~nm) and VIS (533-1020~nm) arms in order to be able to derive Mg and Ca abundances, respectively. It consists of an original sample of 598 UVB spectra (504~stars) and 671 VIS spectra (564 stars). It is worth mentioning that every UVB observation has its counterpart observation in the VIS window, although there are some XSL cool stars (T$_{\rm eff}$~$\textless$~4000~K) from which only the VIS spectral range is available.

\subsubsection{Selection of the best analysed spectra} \label{flags}

The GAUGUIN pipeline was not able to derive reliable individual abundances from certain stellar cases (e.g. atmospheric parameters limitations in the reference synthetic grid, low quality of the observed spectrum, etc.). For instance, we did not consider the derived abundances from the observed spectra that presents S/N~$\textless$~20~pixel$^{-1}$, which are usually unreliable \citep[e.g.][]{smiljanic2014, heiter2015, nissen2018, BIBLIA_Jofre}. This quality cut only affects $\sim$30 UVB spectra. \par

In addition, for each line, GAUGUIN provides a quality indicator that depends on the strength of the spectral line with respect to the noise level, based on an estimate of the detectable limit \citep[upper-limit, for further details see][]{alejandra2022_RVS}. We kept the individual abundances that were above this limit: [X/Fe]~$\textgreater$~[X/Fe]$_{\rm upper-limit}$. We also reject those cases where the measured abundance falls on the synthetic grid border. \par

Thanks to the information provided in \citet{Xshooter_Param}, we trust the XSL stellar parameters for main-sequence stars and those on the giant branch with T$_{\rm eff}$~$\textgreater$~3800~K. The errors in the parameter estimates increase for temperatures lower than 4000~K, and we decided to apply this temperature limit to keep the stars with the most reliable parametrisation (A. Arentsen, priv. comm.). Additionally, as described in Sect.~\ref{Spectra_grids}, the reference synthetic grid is only valid for the analysis of low-rotation stars, since the synthetic spectra are not sensitive to potential line-broadening sources such as stellar rotation and macroturbulence. As a consequence, all stars with T$_{\rm eff}$~$\textgreater$~6500~K were observed to be problematic in the abundance derivation pipeline, showing remarkable discrepancies in the automated fit. In conclusion, we only kept the spectra of stars with 4000~$\textless$~T$_{\rm eff}$~$\textless$~6500~K to minimise the presence of inaccurate abundance values in the final catalogue. \par

\subsubsection{Uncertainties and estimation of the derived stellar abundances} \label{error_estimation}

As explained in Sect.~\ref{GAUGUIN}, for each individual spectrum the GAUGUIN algorithm provides individual chemical abundances for each analysed atomic line of Table~\ref{table:lines}. To estimate the associated internal uncertainties induced by the spectral noise, we implemented 100 different Monte Carlo realisations of each stellar spectrum  by considering flux uncertainties per wavelength pixel. For this purpose, at each Monte Carlo realisation, the original observed stellar flux (before normalisation) was artificially modified per pixel by randomly sampling a Gaussian distribution centred at zero with a standard deviation according to its noise. This produces a total set of 100 values for each individual line abundance. Finally, for each abundance distribution per spectrum, we compute the median and the lower and upper confidence values from the 50th, 16th, and 84th quantiles, respectively. \par

\begin{table*}
\centering
\caption{Abundance uncertainty associated with the stellar atmospheric parameters.}
\begin{tabular}{ccccc|ccc}
\hline
\hline
\\[\dimexpr-\normalbaselineskip+2pt]
 & Cool giant & Cool dwarf & Solar-type & Hot dwarf & [Fe/H] & [Fe/H] & [Fe/H] \\
 & T$_{\rm eff}$ $\sim$ 4700 K & T$_{\rm eff}$ $\sim$ 4700 K & T$_{\rm eff}$ $\sim$ 5800 K & T$_{\rm eff}$ $\sim$ 6200 K & ~$\textless$~-1.0~dex & ~-1.0~-~0.0~dex & ~$\textgreater$~0.0~dex \\
\\[\dimexpr-\normalbaselineskip+2pt]
\hline
\\[\dimexpr-\normalbaselineskip+2pt]
$\Delta$[Mg/Fe]~(dex) & $\pm$~0.022 & $\pm$~0.012 & $\pm$~0.031 & $\pm$~0.044 & $\pm$~0.047 & $\pm$~0.020 & $\pm$~0.017 \\
\\[\dimexpr-\normalbaselineskip+2pt]
$\Delta$[Ca/Fe]~(dex) & $\pm$~0.031 & $\pm$~0.025 & $\pm$~0.057 & $\pm$~0.050 & $\pm$~0.040 & $\pm$~0.041 & $\pm$~0.027 \\
\\[\dimexpr-\normalbaselineskip+2pt]
\hline
\hline
\end{tabular}
\label{table:error_parameters}
\end{table*}

For the final estimation of the star's chemical abundance, we needed to combine the independent abundance measurements of all the available lines per element and consider the cases for which we have repeated spectra. We remind the reader that each analysed spectrum could have up to three magnesium and six calcium measured abundances (we defined two windows at the Ca wing line, see Sect.~\ref{line_list}). First, for a given spectrum, a weighted average of the individual lines per element was calculated following the methodology from \citet{alejandra2022_RVS}, where the mean is weighted by the inverse of the uncertainty of each line (defined as half of the difference between the upper and lower confidence values determined from the 100 Monte Carlo realisations). The final internal associated error of the spectrum is also calculated from the difference of the weighted mean of their upper and lower confidence values. This method allows us to avoid the combined random uncertainties of the different lines. Next, for the stars with more than one observed spectrum, we estimate the final element abundance by averaging the resulting abundances from the previous step. To this purpose, we followed \citet{perdigon2021} \citep[based on the primary study of][]{Vardan2015}, which adopted a weighted mean by the distance from the median in terms of the standard deviation. The final internal error for these cases is proportional to the MAD amongst the repeats ($\Delta$[X/Fe]~=~1.483 × MAD\{[X/Fe]$_{i}$\}). This method allows us to remove the effect of outliers on the final abundance. As an additional internal error analysis, we provide the estimated abundance dispersion for the stars with repeated observed spectra in Appendix~\ref{repeated_spectra_dispersion}. \par

Additionally, we analysed the sensitivity of the derived abundances to the uncertainties on the input XSL stellar atmospheric parameters (T$_{\rm eff}$, log(g), [Fe/H]). For this purpose, without manipulating the stellar spectrum flux, we assigned different sets of the input spectrum parameters within their known uncertainties by implementing 1000 Monte Carlo realisations. We increased the number of realisations for evaluating the combined impact of the three XSL parameters at the same time. This allowed us to derive 1000 different abundance values per spectrum, from which we calculated their standard deviation as an indicate of the induced abundance uncertainty. The results are illustrated in Table~\ref{table:error_parameters} for different stellar types. It can be seen that the averaged abundance dispersion is small, always lower than 0.06~dex, and the derived abundances are more sensitive in hotter and more metal-poor stars, in which the studied lines are weaker. Moreover, the effect is generally stronger in calcium abundances than the magnesium ones. The reason is that Mg lines are strong saturated lines, while for Ca lines we only analysed their wings, which are more sensitive to the abundance variation.


\section{X-shooter catalogue of chemical abundances}\label{results}

Having applied all the quality criteria and the stellar abundance estimation method described in Sect.~\ref{flags_and_errors}, we provide the magnesium and calcium abundances, with their associated internal uncertainty (around $\sim$~0.02~dex, always smaller than 0.1~dex), for 192 and 217 stars, respectively. Due to the limited space, only an extract of the final X-shooter abundance catalogue is reported in Table~\ref{table:data_public}; a full version is available in electronic form. In the public table, the stars are identified by their \emph{Gaia} DR2 \citep{gaia2018} and DR3 IDs \citep{GaiaDR3}, and it contains the median atmospheric parameters from \citet{Xshooter_Param} (T$_{\rm eff}$, log(g), [Fe/H]), the adopted [$\alpha$/Fe] for each case, the number of repeat observed spectra (Nspec), and the averaged S/N of the analysed observed spectra from each X-shooter arm (along with the XSL spectral designations connected to the star); this is combined with the final abundance value, internal error estimate for the star, and its sensitivity to the stellar atmospheric parameters when possible.

\begin{table*}
\centering
\caption{X-shooter catalogue of magnesium and calcium abundances.}
\begin{tabular}{cccccccccccc}
\hline
\hline
\\[\dimexpr-\normalbaselineskip+2pt]
\emph{Gaia} DR3 ID & T$_{\rm eff}$  $^\dagger$ & log(g) $^\dagger$ & [Fe/H]  $^\dagger$ & [$\alpha$/Fe] & [$\alpha$/Fe]$_{\rm source}$ & S/N$_{\rm UVB}$ & [Mg/Fe] & S/N$_{\rm VIS}$ & [Ca/Fe] \\
\\[\dimexpr-\normalbaselineskip+2pt]
\hline
\hline
\\[\dimexpr-\normalbaselineskip+2pt]
2317057722875195904 & 4250.5 & 1.51 & -2.76 & 0.3 & Gaia DR3 & --- & --- & 94.62 & 0.133 \\
\\[\dimexpr-\normalbaselineskip+2pt]
2425521212060616960 & 4349 & 0.67 & -1.82 & 0.4 & Empirical & 47.19 & 0.337 & 102.53 & 0.342 \\
\\[\dimexpr-\normalbaselineskip+2pt]
4685877597088703360 & 4109 & 2.29 & 0.03 & 0.0 & Gaia DR3 & --- & --- & 51 & 0.090 \\
\\[\dimexpr-\normalbaselineskip+2pt]
278890877424832307 & 5164 & 3.77 & -0.7 & 0.28 & Gaia DR3 & 71.59 & 0.288 & 102.12 & 0.223 \\
\\[\dimexpr-\normalbaselineskip+2pt]
253722846670652620 & 4591 & 2.25 & -0.08 & 0.03 & Gaia DR3 & 24.21 & 0.045 & 58.66 & 0.032 \\
\\[\dimexpr-\normalbaselineskip+2pt]
... & ... & ... & ... & ... & ... & ... & ... & ... & ... \\
\\[\dimexpr-\normalbaselineskip+2pt]
... & ... & ... & ... & ... & ... & ... & ... & ... & ... \\
\\[\dimexpr-\normalbaselineskip+2pt]
\hline
\hline
\multicolumn{4}{l}{\scriptsize $\dagger$ Parameters from \citet{Xshooter_Param}}
\end{tabular}
\label{table:data_public}
\end{table*}

\subsection{Magnesium}\label{magnesium}
We first derived the solar magnesium abundance from a very high-S/N HARPS spectrum of Vesta, convolved to the X-shooter UVB arm resolution (R$\sim$10000), to validate the accuracy of our procedure \citep[the same as that of  ][]{perdigon2021}. We obtained a small bias of [Mg/Fe]~=~-0.017~dex, by which the final abundances were calibrated in order to be more consistent with the adopted solar magnesium abundance from the literature. \par

We note that we found a small bias in the derived abundances from the left Mg I 5167.3~$\AA$ line with respect to the two others (see Table~\ref{table:lines}), which is systematically lower with an average difference of $\sim$~0.07~dex (see Appendix~\ref{comparison_lines_abundances} for further details:~Fig.~\ref{Fig:Mg_lines_comparison}). In addition, this line is generally not well-fitted by the described spectrum synthesis algorithm (even for the high-S/N spectrum of Vesta), probably due to the proximity of the strong Fe I 5168.89~$\AA$ line that might be blended and contribute to the derived [Mg/Fe] estimate. The blended nature of this particular Mg I line has been previously reported in some works in the literature \citep[e.g.][]{fuhrmann1997, aldenius2007}. We decided to reject this line in the final abundance analysis.

\begin{figure*}[h]
\centering
\includegraphics[height=50mm, width=\textwidth]{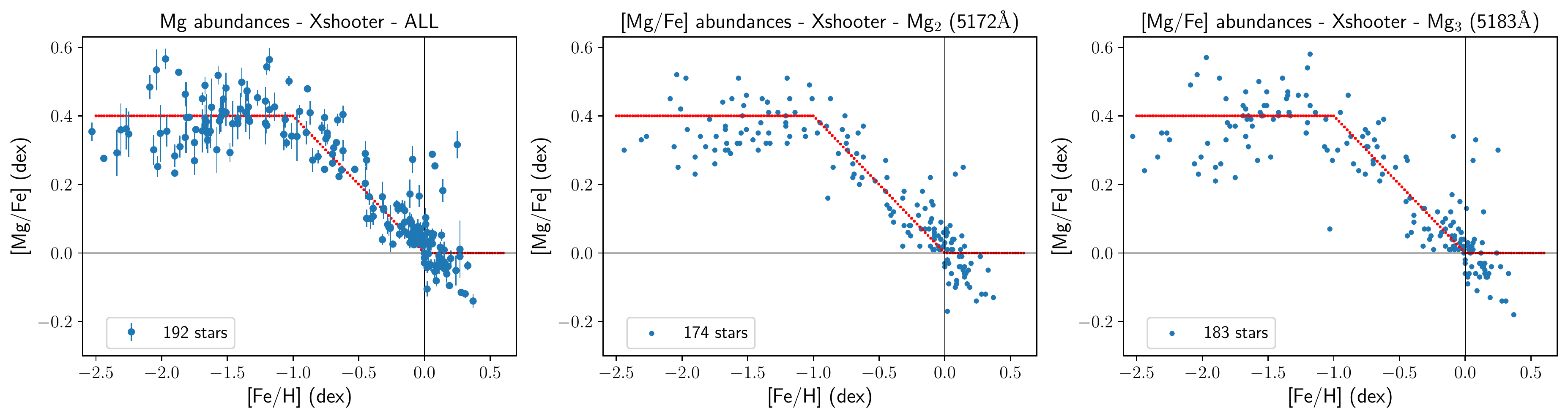}
\caption{Stellar abundance ratios [Mg/Fe] vs. [Fe/H] of the X-shooter catalogue after applying the optimal method. From left to the right, we show the derived abundances from the considered Mg Ib triplet lines with the estimated internal uncertainties as vertical error bars and each individual line separately: 5172.7~$\AA$ and 5183.6~$\AA$. The red line reproduces the empirical [$\alpha$/Fe] approach adopted in the analysis when no reliable values from \emph{Gaia} DR3 are found.}
\label{Fig:Mg_abundances}
\end{figure*}

Figure~\ref{Fig:Mg_abundances} illustrates the final stellar abundance ratios [Mg/Fe] relatively to the stellar metallicity [Fe/H] for the best analysed XSL sample, along with their estimated internal uncertainties from the combined analysis of the two last Mg Ib triplet lines (shown in Table~\ref{table:lines}). The stars ($\sim$~55\% are giants) cover a metallicity domain from $\sim$ -2.5~dex to +0.4~dex and seem to reproduce a flat trend in the metal-poor regime ([Fe/H]~$\lesssim$~-1.0~dex) followed by a decrease in [Mg/Fe] even at supersolar metallicities ([Fe/H]~$\textgreater$~0.0~dex). This behaviour is in perfect agreement with Galactic chemical evolution models \citep[][]{chiappini1997, kobayashi2006, romano2010, spitoni2020, Palla2022}. The sample seems to be dominated by Galactic disc stars, although metal-poor stars ([Fe/H]~$\textless$~-1.0~dex) are mostly halo stars. As described in \citet{XshooterDR2}, XSL DR2 stars are located in star clusters, in the field, in the Galactic bulge, and in the Magellanic Clouds (the reference stellar catalogues can be found in their Table A.1. with the respective target selection). A complementary dynamical selection of the different Galactic stellar populations is beyond the scope of this paper.

\subsubsection{Comparison with \emph{Gaia} DR3}\label{magnesium_Gaia}

In this section, we compare our derived [Mg/Fe] abundances from the XSL sample (R~$\sim$~10000) and those recently provided by \emph{Gaia} DR3 \citep{alejandra2022_RVS} from RVS observations (R~$\sim$~11500), where the same analysis procedure was implemented. However, due to the limited spectral range of the RVS (8460-8700~$\AA$), \emph{Gaia} DR3 [Mg/Fe] abundances were only derived from the weak Mg I line 8476.02~$\AA$.  \par

\begin{figure*}[h]
\centering
\includegraphics[height=50mm, width=\textwidth]{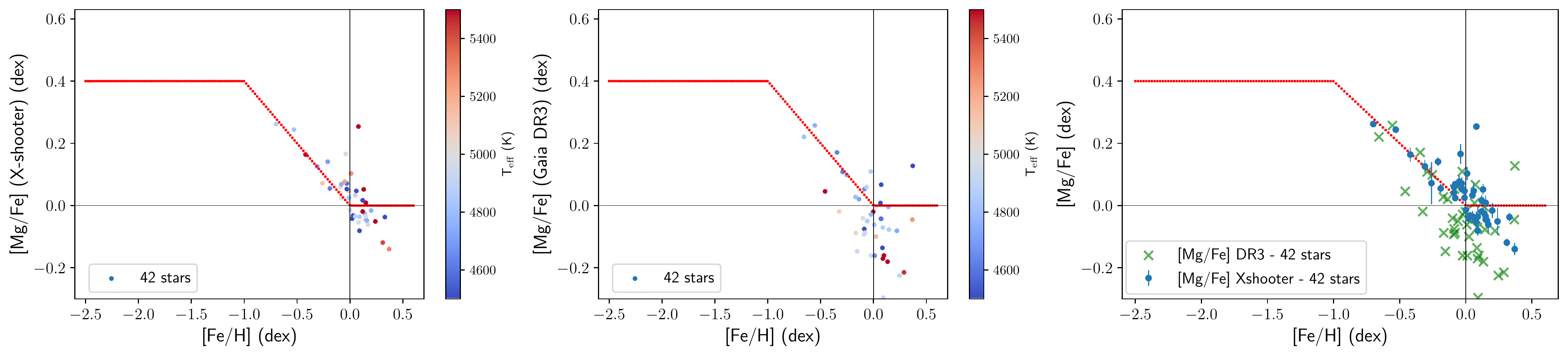}
\includegraphics[height=50mm, width=1.01\textwidth]{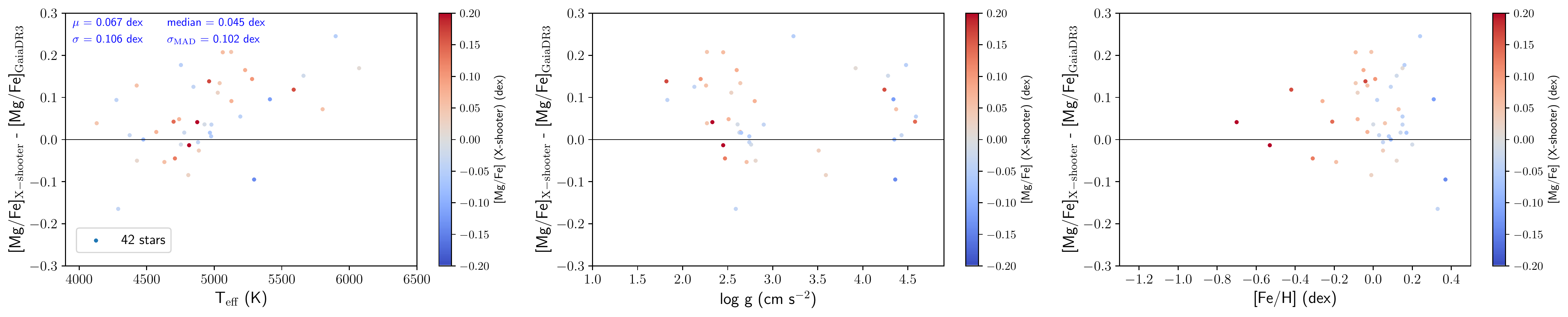}
\caption{Comparison between derived [Mg/Fe] abundances in this work and those provided from \emph{Gaia} DR3 data. \textit{Top row:} Abundance ratios [Mg/Fe] vs. [Fe/H] from X-shooter (\textit{top-left}), \emph{Gaia} DR3 (\textit{top-middle}), and both catalogues together  (\textit{top-right}, \emph{Gaia} DR3 data are represented by green crosses), colour-coded by the stellar effective temperature in the first two panels. The red line is the same as Fig.~\ref{Fig:Mg_abundances}. \textit{Bottom row:} Stellar abundance difference between catalogues as a function of the effective temperature (\textit{bottom-left}), surface gravity (\textit{bottom-middle}), and metallicity (\textit{bottom-right}), colour-coded by the measured [Mg/Fe] abundance. The offset estimates are the same as those described in Fig.~\ref{Fig:Parameter_comparison}.}
\label{Fig:Mg_comparison_Gaia}
\end{figure*}

For this purpose, in addition to the cross-match described in Sect.~\ref{Parameter_Validation}, we selected the stars from the \emph{Gaia} DR3 dataset that fullfil the two \emph{Gaia} reliability flags (XUpLim, XUncer) defined for each individual abundance in \citet{alejandra2022_RVS}. These abundance flags are based on the detectable limit and the reliability of the associated uncertainties from each analysed line, which are imposed to be equal to zero in order to ensure the best abundance quality. Moreover, the individual abundance calibrations indicated by \citet{alejandra2022_RVS} were also applied for an appropriate comparison with our studied XSL sample. \par

The cross-match leads to a total of 42 stars in common with good quality  \emph{Gaia} DR3 [Mg/Fe] abundances. Figure~\ref{Fig:Mg_comparison_Gaia} shows the [Mg/Fe]-[Fe/H] plane for both samples separately (XSL in the top-left panel, \emph{Gaia} DR3 in the top-middle panel), colour-coded with the star's effective temperature, and overplotted together in the top right panel. The empirical [$\alpha$/Fe] approach adopted in the analysis when no reliable values from \emph{Gaia} DR3 are found (see Sect.~\ref{flags}) is also given as a visual reference to aid abundance comparison. Additionally, the abundance difference between both catalogues is also represented as a function of the stellar effective temperature (bottom left panel), surface gravity (bottom middle panel), and metallicity (bottom right panel); this is colour-coded according to the measured XSL [Mg/Fe] abundance. In general, we observe a good agreement between both [Mg/Fe] abundance datasets, reproducing a similar trend with metallicity, although our derived abundances seem to present slightly higher values compared to the \emph{Gaia} ones (with a median offset of $\sim$~0.05~dex), showing an apparent increase with temperature and no clear systematic dependences on the other stellar parameters. \par 

The reported abundance differences could come from the fact that the analysed Mg I lines are not the same in both studies. As mentioned before,  \emph{Gaia} DR3 analysis was centred on the weak Mg I line 8476.02~$\AA$, which strength is significantly lower in comparison with our strong lines of the Mg Ib triplet. As analysed in detail in \citet{SantosPeral2020}, the normalisation has a dependence on the line intensity that could remain in the abundance results. Strong lines span larger wavelength domains, farther from the spectra noise level, and present a more cumulative quantity of information on the abundance, which may lead to a more precise determination.

\subsubsection{Comparison with AMBRE:HARPS} \label{magnesium_SantosPeral}

As discussed above, we followed the same applied methodology as \citet{SantosPeral2020} for deriving stellar abundances, where the normalisation procedure for different stellar types and each analysed optical Mg line was optimised separately. They obtained precise and accurate [Mg/Fe] abundances for a total of 1066 solar neighbourhood stars (\citealp{SantosPeral2020, SantosPeral2021}; both compiled in \citealp{Palla2022}) observed at high spectral resolution (R~$\sim$~115000) by the HARPS spectrograph \citep{mayor2003}. The whole AMBRE:HARPS stellar sample is fully described in \citet{DePascale2014}. \par

\begin{figure}
\centering
\hspace*{-0.5cm}
\includegraphics[height=65mm, width=0.45\textwidth]{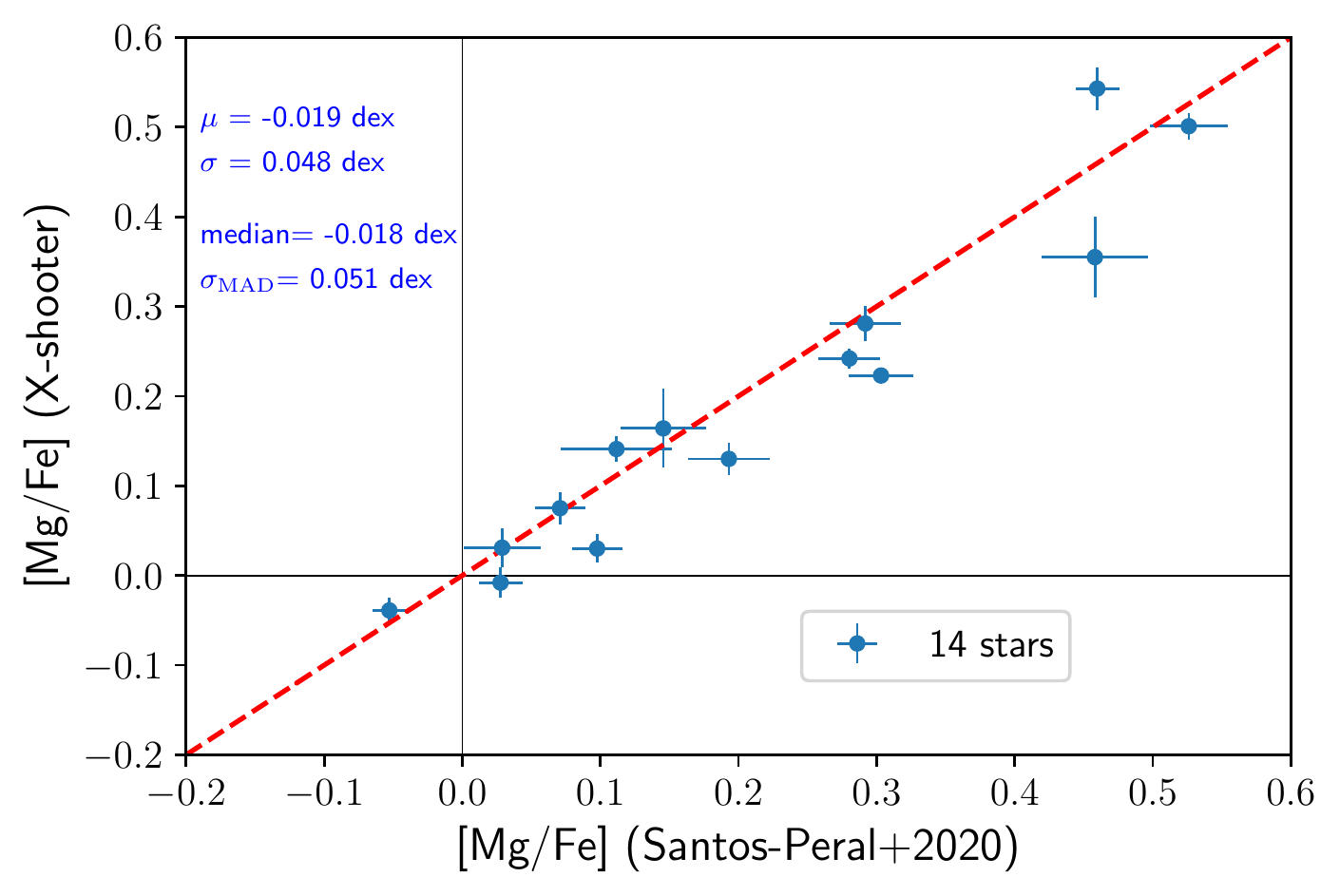}
\caption{Direct comparison of stars in common between the derived stellar abundance ratio [Mg/Fe] in this work and the precise abundances from the previous work of \citet{SantosPeral2020} for the AMBRE:HARPS sample. The red dashed line reproduces the linear relation (y~=~x).}
\label{Fig:Mg_Xshooter_SantosPeral}
\end{figure}

Therefore, as an additional check of our derived XSL [Mg/Fe] abundances, we carried out a cross-match with the AMBRE:HARPS dataset, from which we only found 14 stars in common to compare. Figure~\ref{Fig:Mg_Xshooter_SantosPeral} illustrates the direct [Mg/Fe] abundance comparison with the associated internal uncertainties from both studies. It shows a remarkable agreement with just a small averaged difference of -0.019~dex, where some of the stars present almost identical values.

\subsection{Calcium}\label{calcium}

\begin{figure*}
\centering
\includegraphics[height=50mm, width=\textwidth]{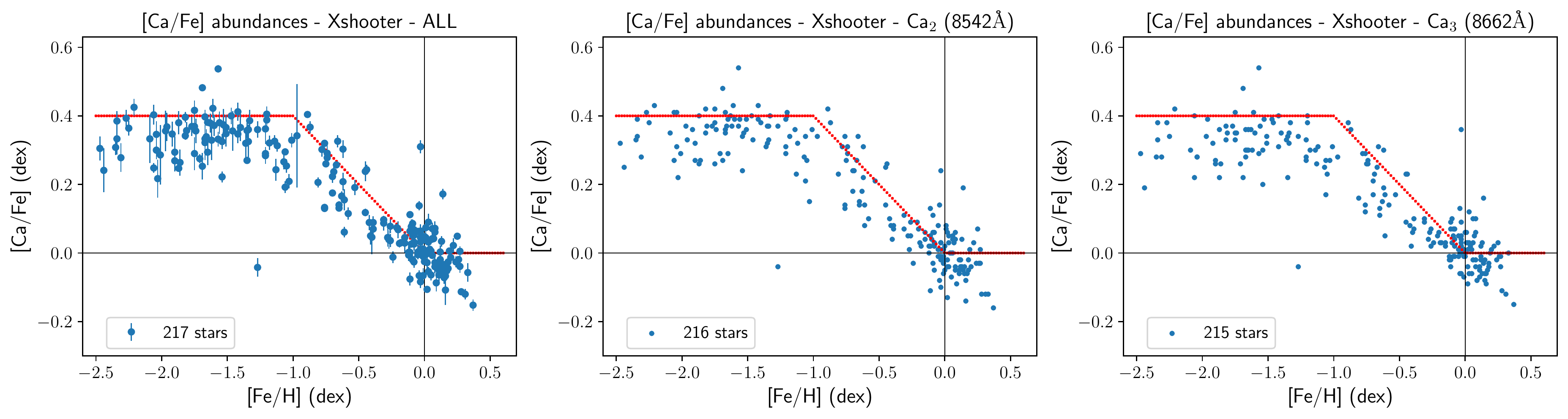}
\caption{Same as Fig.~\ref{Fig:Mg_abundances}, but for the final [Ca/Fe] abundance X-shooter catalogue: considering both wings of the Ca II lines (\textit{left panel}) and each individual line separately: 8542.09~$\AA$ and 8662.14~$\AA$.}
\label{Fig:Ca_abundances}
\end{figure*}

Similarly to the reported behaviour of the first Mg Ib triplet line (5167.3~$\AA$, Sect.~\ref{magnesium}), we observed an averaged systematic bias of $\sim$~0.05~dex (see~Fig.~\ref{Fig:Ca_lines_comparison} in Appendix~\ref{comparison_lines_abundances}) and much more dispersed derived abundances from the left Ca II 8498.02~$\AA$ line in comparison with the two last calcium lines (see Table~\ref{table:lines}). The difference is even larger when we isolate the left wing of the line ($\sim$~0.12~dex). The reason behind is the presence of the strong titanium line, Ti I 8468.50~$\AA$, an $\alpha$ element that may affect the normalisation of the calcium line, although the additional presence of close Fe I lines (such as the Fe I line~8496.99~$\AA$) could be also responsible \citep[as already discussed in previous studies of the Ca II triplet region; e.g.][]{mallik1997, chmielewski2000, cenarro2001}. We also decided to reject this line in the final [Ca/Fe] abundance analysis. \par

\begin{figure*}
\centering
\includegraphics[height=50mm, width=\textwidth]{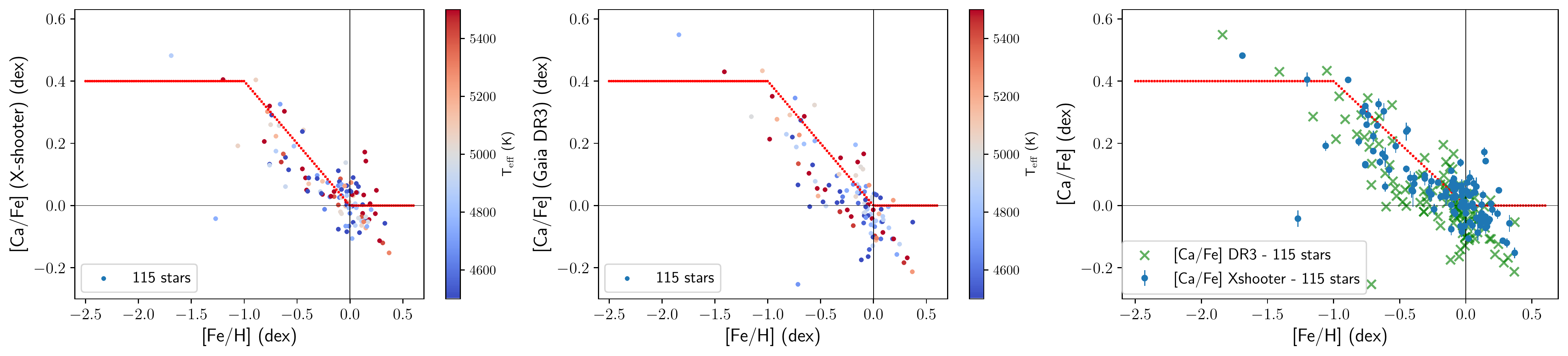}
\includegraphics[height=50mm, width=1.01\textwidth]{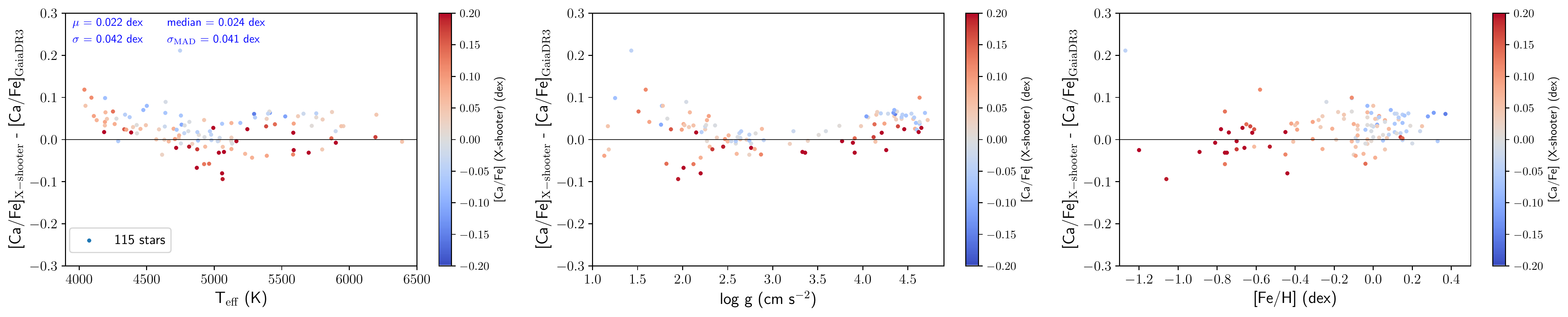}
 \caption{Same as Fig.~\ref{Fig:Mg_comparison_Gaia}, but for the [Ca/Fe] abundances in common with good quality \emph{Gaia} DR3 data.}
\label{Fig:Ca_comparison_Gaia}
\end{figure*}

Figure~\ref{Fig:Ca_abundances} illustrates the final [Ca/Fe] abundances as a function of the stellar metallicity [Fe/H] for the best analysed sample ($\sim$~60\% of giants), together with their associated internal uncertainties. As shown for the [Mg/Fe] catalogue in Sect.~\ref{magnesium}, the [Ca/Fe] sample covers the same metallicity domain, and reproduces the described flat and decreasing trend in the metal-poor and metal-rich regimes, respectively. Therefore, we can confirm from our results the $\alpha$ nature of both elements, which is in agreement with the observed and expected global [$\alpha$/Fe] trend.

\subsubsection{Comparison with \emph{Gaia} DR3}\label{calcium_Gaia}

We remind the reader that we analysed the same Ca II lines  (Table~\ref{table:lines}) as those considered in \citet{alejandra2022_RVS} in this work, using their 5D synthetic spectra grid (Sect.~\ref{Spectra_grids}) and applying the same procedure for the individual [Ca/Fe] chemical abundance determination in the \emph{Gaia} DR3 catalogue. Moreover, it is worth mentioning that calcium is one of the most robust elements determined in \emph{Gaia}, where wavelength coverage is centred around the Ca II IR triplet lines that dominate the spectral analysis of the RVS. For these reasons, the comparison with \emph{Gaia} DR3 [Ca/Fe] abundances is crucial to calibrate and validate the accuracy of our measurements from the X-shooter spectra. \par

Following the same described quality criteria for the [Mg/Fe] analysis (Sect.~\ref{magnesium_Gaia}), we found a larger number of stars in common with reliable [Ca/Fe] abundances from the \emph{Gaia} DR3 data: 115 stars. Figure~\ref{Fig:Ca_comparison_Gaia} shows an excellent agreement between both [Ca/Fe] abundance catalogues, reproducing the same trend in the [Ca/Fe]-[Fe/H] plane and tracing an almost linear correlation with a very small average bias of 0.022~dex, which is slightly larger for lower [Ca/Fe] abundances that correspond to the more metal-rich stars. However, it is interesting to remark the described curve as a function of the surface gravity, where an offset is observed in the giant and dwarf star regimes. A gravity dependence of the Ca II IR triplet lines has previously been reported \citep[e.g.][]{Cenarro2002}, and its effect in the analysed calcium line wings cannot be completely discarded in both the \emph{Gaia} and the X-shooter determinations. The strong consistency of our measurements with \emph{Gaia} DR3 abundances, which were carefully calibrated as a function of the analysed stellar type \citep{alejandra2022_RVS}, allow us to certify the accuracy of the derived [Ca/Fe] abundances from the XSL sample with no need for any further calibration.

\subsection{Stars with both [Mg/Fe] and [Ca/Fe] abundances} \label{magnesium_and_calcium}

In this section, we summarise the X-shooter stellar sub-sample of 174 stars from which we provide reliable [Mg/Fe] and [Ca/Fe] abundances. \par

\begin{figure*}
\centering
\hspace*{-0.3cm}
\includegraphics[height=50mm, width=1\textwidth]{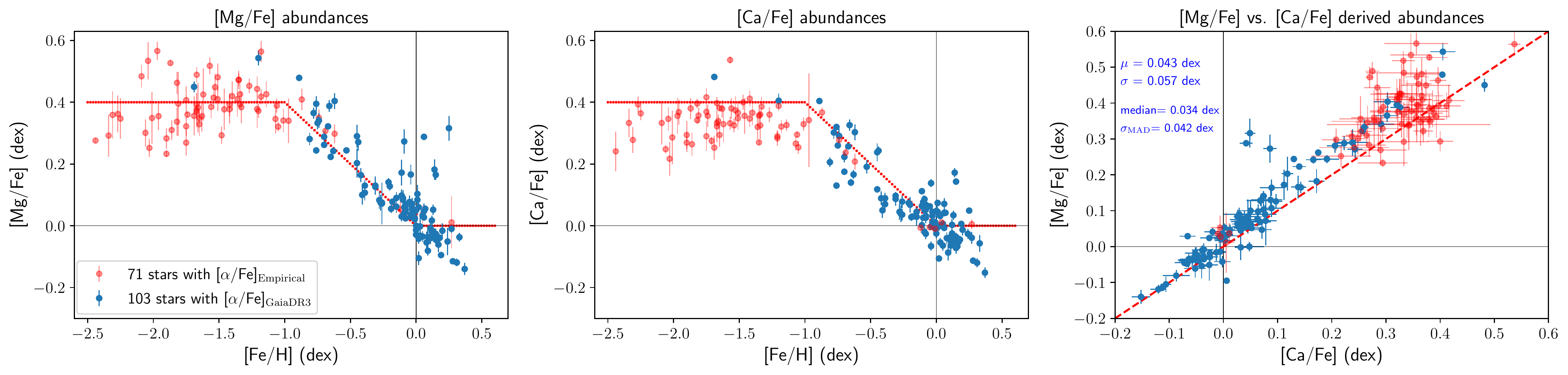}
\caption{Comparison of the derived stellar abundance ratios from the final X-shooter stellar sample with reliable [Mg/Fe] and [Ca/Fe] abundances, colour-coded by the origin of the adopted input [$\alpha$/Fe] parameter: from \emph{Gaia} DR3 (blue) or empirical approach (red). \textit{Left and middle panels}: [Mg/Fe] and [Ca/Fe] as a function of [Fe/H], respectively. \textit{Right panel}: direct comparison [Mg/Fe] vs. [Ca/Fe].}
\label{Fig:MgvsCa_Xshooter}
\end{figure*}

\begin{figure*}
\centering
\hspace*{-0.3cm}
\includegraphics[height=50mm, width=1.05\textwidth]{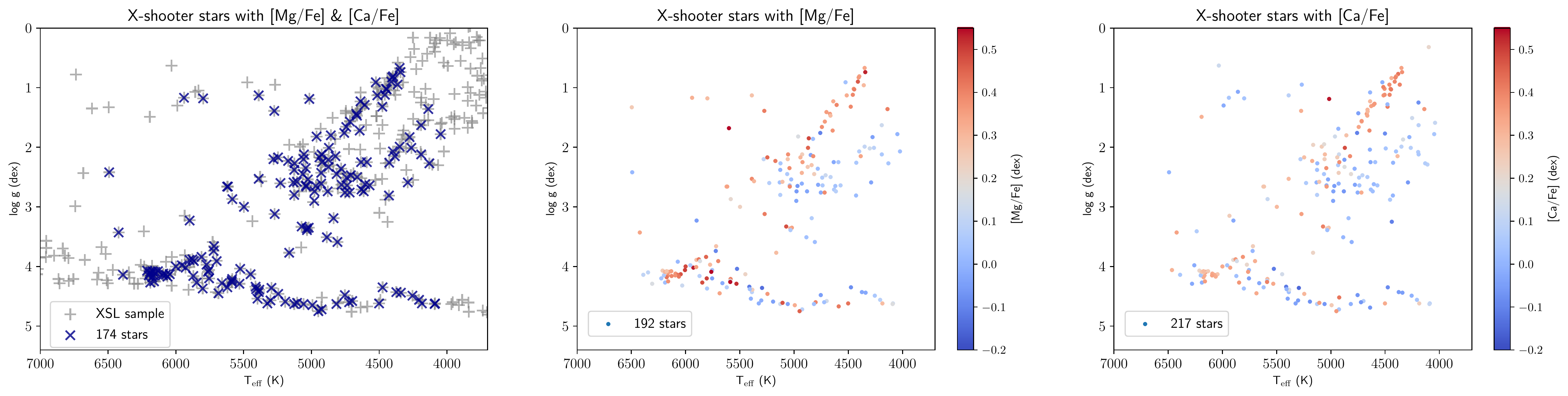}
\caption{HR diagrams of the final X-shooter stellar sample with reliable [Mg/Fe] and [Ca/Fe] abundance measurements. \textit{Left panel}: stars chemically characterised with [Mg/Fe] and [Ca/Fe] abundances (blue crosses) over the whole analysed X-shooter sample in this work (grey crosses). \textit{Middle and right panels}: stars with reliable [Mg/Fe] and [Ca/Fe] estimates, colour-coded by the element abundance respectively.}
\label{Fig:HR_abundances}
\end{figure*}

The results for each element are presented separately and comparatively in Fig.~\ref{Fig:MgvsCa_Xshooter}. As already discussed in Sect.~\ref{magnesium} and Sect.~\ref{calcium}, both stellar abundance ratios [X/Fe] describe the same trend as a function of the metallicity [Fe/H] (left and middle panels), which is typical for $\alpha$-element evolution. In addition, we measured relatively higher magnesium abundances in comparison with calcium values (with a systematic difference at $\sim$~0.034~dex, right panel), which is more easily noticeable for metal-poor stars ([Fe/H]~$\lesssim$~-1.0~dex) where the [Mg/Fe] abundance shows a larger dispersion. This behaviour has been previously reported in the literature from both observational \citep[e.g. element abundance from APOGEE data,][]{zasowski2019} and theoretical points of view \citep[stellar nucleosynthesis yields: e.g.][]{francois2004, kobayashi2006}. [Ca/Fe] tends to be under-abundant, relatively to other $\alpha$ elements, for more massive and more metal-rich supernovae. In addition, the discussed behaviour could be explained by a possible mass or metallicity dependence in the Ca yields of core-collapse supernovae (Type II SNe), which is commonly adopted for oxygen \citep[e.g.][]{GoswamiPrantzos2000}, although these dependences have not yet been found in the literature. \par

The covered parameters in the HR diagram by the selected stars with measurements in both elements are shown in Fig.~\ref{Fig:HR_abundances}. The observed colour gradient in the abundances for giant stars is a result of the mentioned decreasing trend with the stellar metallicity. We satisfactorily obtained chemical abundances for a wide variety of stellar types (dwarfs and giants), well distributed in the effective temperature (4000~$\textless$~T$_{\rm eff}$~$\textless$~6500~K) and the metallicity (-2.5~$\textless$~[Fe/H]~$\textless$~+0.4~dex) range. The completeness of the provided catalogue in the atmospheric parameter and chemical space seems to be ideal for improving the development of evolutionary stellar population synthesis models in the near future for the analysis of external galaxies abundances.





\section{Conclusions} \label{conclusions}

In this paper, we give a detailed spectroscopic analysis of the Mg and Ca abundance estimation for the XSL Data Release~2 (medium resolution library, R~$\sim$~10000). We obtained accurate [Mg/Fe] and [Ca/Fe] abundances for 192 and 217 stars, respectively. 174 of these stars have measurements of both elements, with a wide coverage of the parameter space ($\sim$~56\% giants; 4000~$\textless$~T$_{\rm eff}$~$\textless$~6500~K; -2.5~$\textless$~[Fe/H]~$\textless$~+0.4~dex). \par 

We used the UVB and VIS arms of the instrument and derived abundances from individual spectral lines with the spectrum synthesis code GAUGUIN. 
We also performed additional parameter validations with the \emph{Gaia} DR3 (R~$\sim$~11500) and the AMBRE Project (R~$\sim$~47000-115000) catalogues, finding an excellent agreement. \par

The final derived stellar abundance ratios [X/Fe], relative to the star's metallicity [Fe/H], present a flat behaviour in the metal-poor regime ([Fe/H]~$\lesssim$~-1.0~dex) followed by a decreasing trend even at supersolar metallicities ([Fe/H]~$\textgreater$~0.0~dex), as observed and expected for $\alpha$-nature elements by Galactic chemical evolution models. In addition, we remark the close agreement found with the Mg and Ca abundance estimates from the AMBRE:HARPS study by \citet{SantosPeral2020} and the recent \emph{Gaia} DR3 analysis of RVS observations \citep{alejandra2022_RVS}, which both implemented the same procedure. \par


The provided XSL abundance catalogue, which has an excellent parameter coverage for intermediate and old stellar populations and a wide metallicity range, follows the abundance pattern of the Galaxy. Therefore, stellar population models based on the XSL will provide predictions that follow the same pattern \citep{Verro2022}. However, the abundance ratios determined here are required for building up models with varying Mg and Ca abundances. Differential corrections obtained by varying the abundance ratios of these elements can be used to correct the XSL stars to compute semi-empirical spectra with varying levels of Mg and Ca abundance ratios and, consequently, the stellar population models that implement these stars. These models will allow us to study small galaxies with nearly solar-scaled abundances in the sub-solar and metal-poor metallicity regimes and massive galaxies with enhanced [$\alpha$/Fe] in the high-metallicity regime. In conclusion, this will significantly contribute to the quality of research of external galaxies, with a special focus placed on current and next-generation of field spectrographs, such as MEGARA \citep{GildePaz2018}, WEAVE \citep{weave}, or 4MOST \citep{4MOST}.

\begin{acknowledgements}
The authors thank A. Recio-Blanco, P. de Laverny, C. Ordenovic and M. A. \'{A}lvarez for having developed the GAUGUIN method within the \emph{Gaia}/GSP-spec context and having shared it with us. For any questions related to this algorithm, please contact directly A. Recio-Blanco and P. de Laverny. We thank the referee for the careful reading of the manuscript,  his/her constructive comments have significantly improved the quality of the paper. P.S.P and P.S.B acknowledges financial support by the Spanish Ministry of Science and Innovation through the research grant PID2019-107427GB-C31. P.S.P also acknowledges financial support by the European Union - NextGenerationEU under a Margarita Salas contract. A.V. acknowledges support from grant PID2019-107427GB-C32 from the Spanish Ministry of Science, Innovation and Universities MCIU. This work has also been supported through the IAC project TRACES, which is partially supported through the state budget and the regional budget of the Consejer{\'{i}}a de Econom{\'{i}}a, Industria, Comercio y Conocimiento of the Canary Islands Autonomous Community. A.V. also acknowledges support from the ACIISI, Consejer{\'{i}}a de Econom{\'{i}}a, Conocimiento y Empleo del Gobierno de Canarias and the European Regional Development Fund (ERDF) under grant with reference ProID2021010079. P.A.P acknowledges the financial support from the Centre national d’études spatiales (CNES). This work has made use of data from the European Space Agency (ESA) mission {\it Gaia} (\url{https://www.cosmos.esa.int/gaia}), processed by the {\it Gaia} Data Processing and Analysis Consortium (DPAC, \url{https://www.cosmos.esa.int/web/gaia/dpac/consortium}). Funding for the DPAC has been provided by national institutions, in particular the institutions participating in the {\it Gaia} Multilateral Agreement.  Most of the calculations have been performed with the high-performance computing facility SIGAMM, hosted by the Observatoire de la Côte d'Azur (OCA).
\end{acknowledgements}



\bibliographystyle{aa}  
\bibliography{Santos-Peral} 

\begin{appendix} 

\section{Abundance dispersion from repeated observed spectra} \label{repeated_spectra_dispersion}

\begin{figure}[h]
\centering
\hspace*{-0.5cm}
\includegraphics[height=70mm, width=0.5\textwidth]{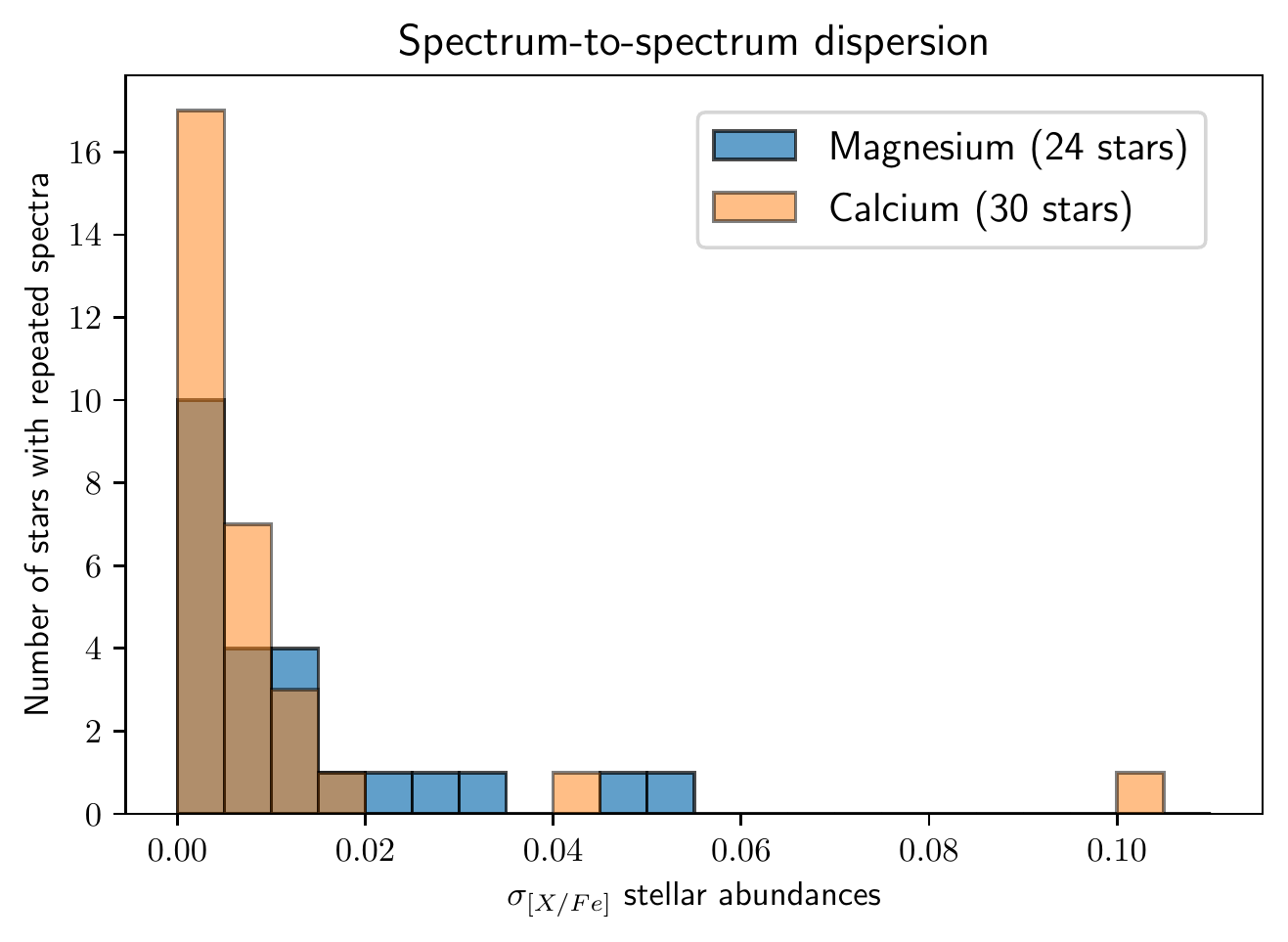}
\caption{Estimated dispersion on derived [Mg/Fe] (blue) and [Ca/Fe] (orange) abundances of the stars in the final catalogue with repeated observations ($\geq$~2 spectra).}
\label{Fig:MgCa_histogram_dispersion}
\end{figure}

For some analysed stellar cases, the X-shooter spectral catalogue contains repeated observations. The amount ($\sim$~25-30~stars) is not significant in the final sample, which has good quality derived abundances. The estimated dispersion on the magnesium and calcium abundance measurements from the different spectra of the same star are shown in Fig.~\ref{Fig:MgCa_histogram_dispersion}. The spectrum-to-spectrum dispersion is generally lower than 0.02~dex, with an average dispersion on [Mg/Fe] and [Ca/Fe] of 0.013 and 0.010~dex, respectively.

\section{Line-to-line abundance comparison} \label{comparison_lines_abundances}

\begin{figure}[h]
\centering
\includegraphics[width=0.9\textheight, width=0.48\textwidth]{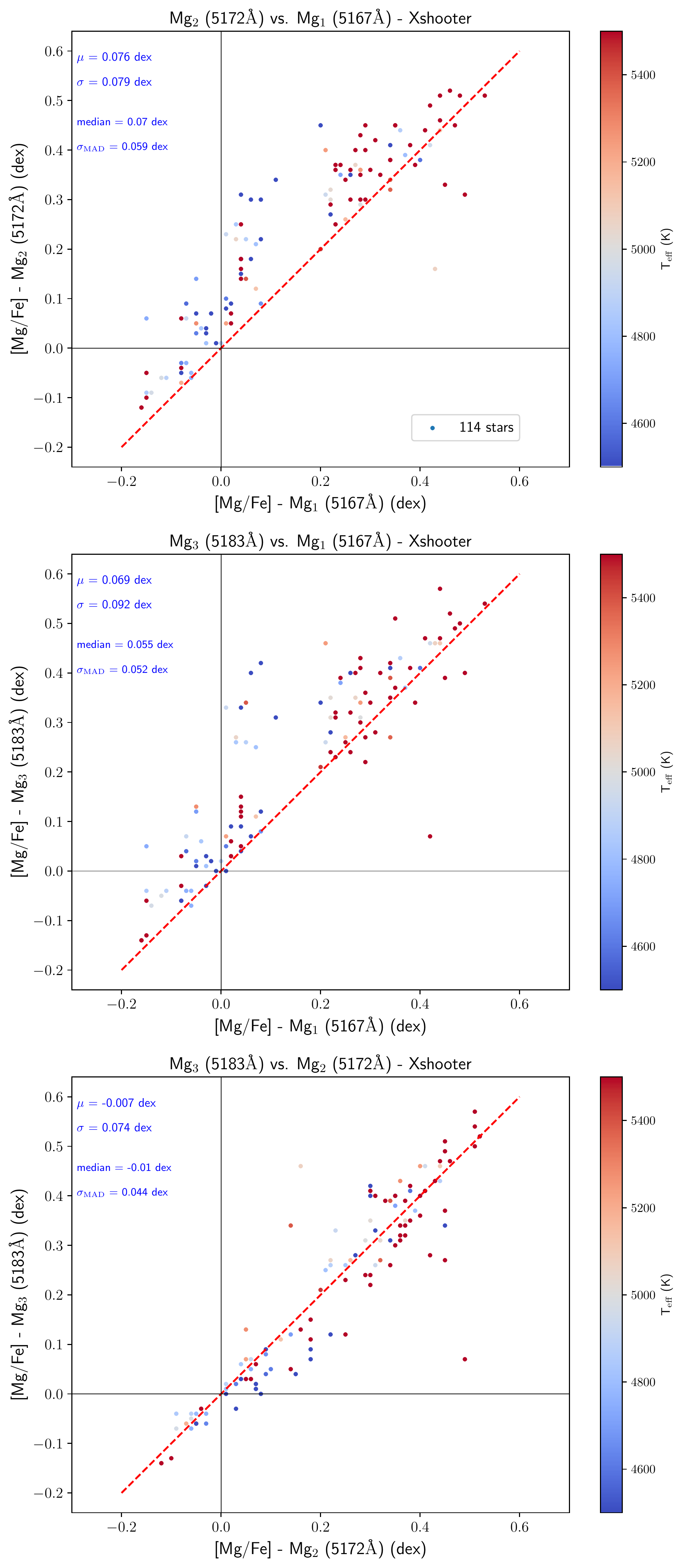}
\caption{Comparison, colour-coded according to stellar effective temperature, between the derived [Mg/Fe] abundances from the individual analysed spectral lines: Mg$_{\rm 1}$ (5167.3~$\AA$), Mg$_{\rm 2}$ (5172.7~$\AA$), and Mg$_{\rm 3}$ (5183.6~$\AA$). \textit{Top:} [Mg$_{\rm 2}$/Fe] vs. [Mg$_{\rm 1}$/Fe]. \textit{Middle:} [Mg$_{\rm 3}$/Fe] vs. [Mg$_{\rm 1}$/Fe]. \textit{Bottom:} [Mg$_{\rm 3}$/Fe] vs. [Mg$_{\rm 2}$/Fe]. The red dashed line reproduces the linear relation (y~=~x). The offset estimates are the same as described in Fig.~\ref{Fig:Parameter_comparison}.}
\label{Fig:Mg_lines_comparison}
\end{figure}

We evaluated possible abundance internal biases in our method by comparing the individual [X/Fe] estimates from each spectral line individually. Figure~\ref{Fig:Mg_lines_comparison} shows the direct [Mg/Fe] line-to-line abundance comparison among the three analysed Mg Ib triplet lines (5167.3, 5172.7, and 5183.6~$\AA$), colour-coded with the star’s effective temperature. In the first two panels, we observe a clear bias with respect to the Mg I 5167.3~$\AA$ line without a dependence on the stellar type. We find a similar abundance offset of $\sim$~0.07~dex for both Mg 5172.7 and 5183.6~$\AA$ lines. However, we measured compatible abundance values from these two lines (bottom panel), almost reproducing a perfect linear relation. \par 

\begin{figure*}[h]
\centering
\includegraphics[height=130mm, width=0.9\textwidth]{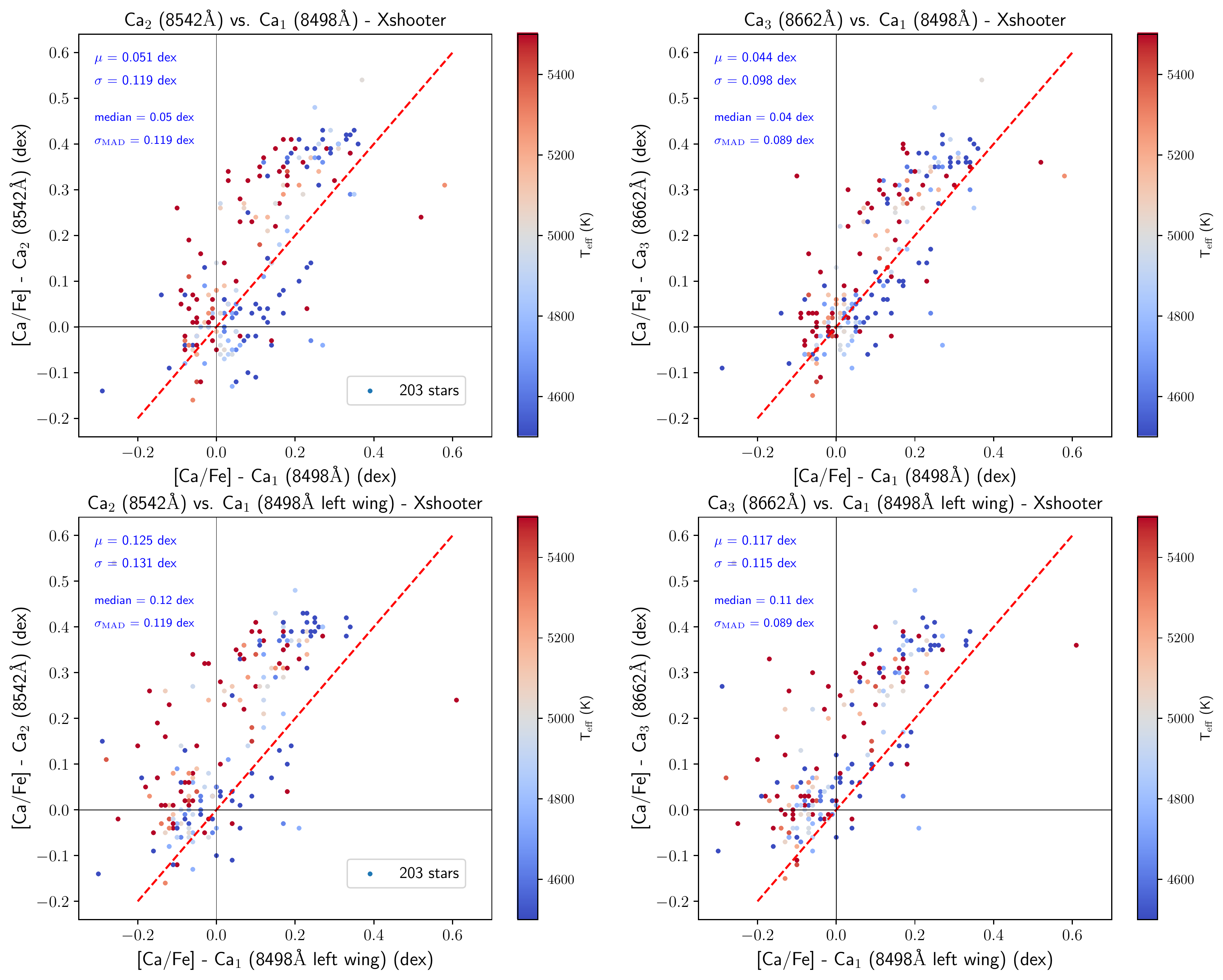}
\hspace*{-0.8cm}
\includegraphics[height=70mm, width=0.45\textwidth]{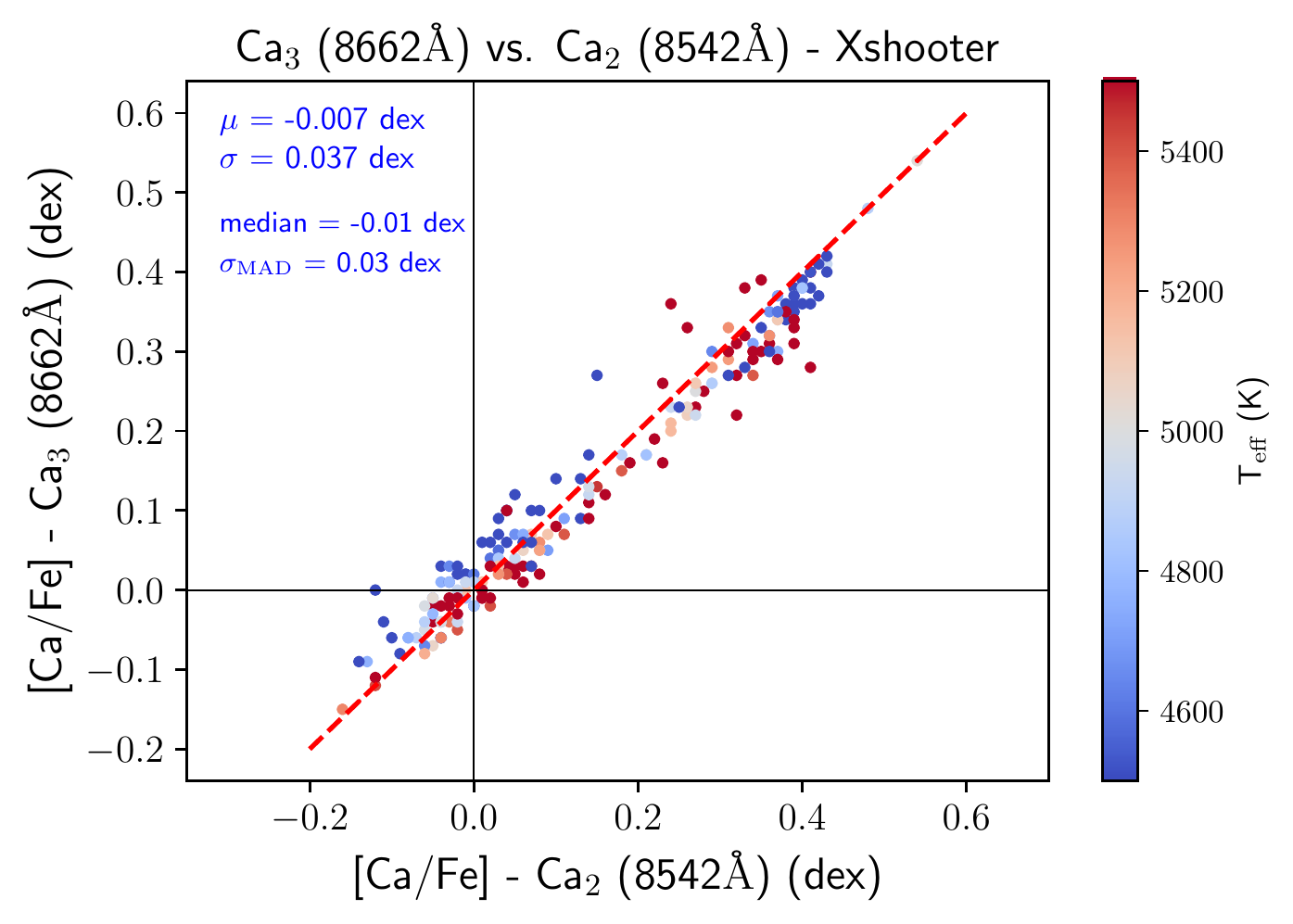}
\caption{Same as Fig.~\ref{Fig:Mg_lines_comparison}, but for the derived [Ca/Fe] abundances from Ca$_{\rm 1}$ (8498.02~$\AA$), Ca$_{\rm 2}$ (8542.09~$\AA$), and Ca$_{\rm 3}$ (8662.14~$\AA$). \textit{Top row:} [Ca$_{\rm 2}$/Fe] vs. [Ca$_{\rm 1}$/Fe] (\textit{top-left}) and [Ca$_{\rm 3}$/Fe] vs. [Ca$_{\rm 1}$/Fe] (\textit{top-right}). \textit{Middle row:} Same when only comparing with the left wing of Ca$_{\rm 1}$. \textit{Bottom row:} [Ca$_{\rm 3}$/Fe] vs. [Ca$_{\rm 2}$/Fe].}
\label{Fig:Ca_lines_comparison}
\end{figure*}

On the other hand, the [Ca/Fe] line-to-line abundance is illustrated in Fig.~\ref{Fig:Ca_lines_comparison} for the considered Ca II IR triplet: 8498.02, 8542.09, and 8662.14~$\AA$. We report a significant scatter of the Ca~II 8498.02~$\AA$ line in comparison with the other two lines (top left and top right panels), with an average offset of $\sim$~0.05~dex. These differences increase up to $\sim$~0.12~dex when we only compare with the left wing of the line (middle row panels). Furthermore, we observe a remarkably good agreement between the independent abundance estimates from the last two calcium lines.

\end{appendix}

\end{document}